\documentclass[usenatbib]{mnras} 

\usepackage{newtxtext,newtxmath}

\usepackage[T1]{fontenc}

\DeclareRobustCommand{\VAN}[3]{#2}
\let\VANthebibliography\thebibliography
\def\thebibliography{\DeclareRobustCommand{\VAN}[3]{##3}\VANthebibliography}

\usepackage{graphicx}	
\usepackage{amsmath}	
\usepackage{hyperref}   
\usepackage{xcolor}
\usepackage{rotating}
\usepackage{orcidlink}

\title[Ly$\alpha$ escape in JELS-MUSE]{Constraining reionization-era Ly$\alpha$ escape with JELS-MUSE: a highly complete H$\alpha$-selected sample at $z\sim6.1$} 

\author[A. L. Patrick et al.]{
A. L. Patrick\,\orcidlink{0000-0003-0645-6853}\,$^{1}$\thanks{E-mail: abigail.patrick@ed.ac.uk (ALP)},
K. J. Duncan\,\orcidlink{0000-0001-6889-8388}\,$^{1}$,
Z. Li\,\orcidlink{0000-0001-7373-3115}\,$^{2}$,
C. A. Pirie\,\orcidlink{0009-0003-5303-6920}\,$^{1}$,
A. M. Swinbank\,\orcidlink{0000-0003-1192-5837}\,$^{2}$, 
L. C. Keating\,\orcidlink{0000-0001-5211-1958}\,$^{1}$,\newauthor
S. R. Flury\,\orcidlink{0000-0002-0159-2613}\,$^{1}$,
R. Begley\,\orcidlink{0000-0003-0629-8074}\,$^{1,3}$,
P. N. Best\,\orcidlink{0000-0001-5081-4801}\,$^{1}$,
M. Brinch\,\orcidlink{0000-0002-0245-6365}\,$^{4,5}$,
A. C. Carnall\,\orcidlink{0000-0002-1482-5818}\,$^{1}$,
F. Cullen\,\orcidlink{0000-0002-3736-476X}\,$^{1}$,
J. S. Dunlop\,\orcidlink{0000-0002-1404-5950}\,$^{1}$,\newauthor
E. Ibar\,\orcidlink{0009-0008-9801-2224}\,$^{4,5}$,
J. Matthee\,\orcidlink{0000-0003-2871-127X}\,$^{6}$,
D. J. McLeod\,\orcidlink{0000-0003-4368-3326}\,$^{1}$, 
A. Puglisi\,\orcidlink{0000-0001-9369-1805}\,$^{7}$,
H. M. O. Stephenson\,\orcidlink{0000-0002-0777-1591}\,$^{8}$
and J. P. Stott\,\orcidlink{0000-0002-1679-9983}\,$^{8}$
\\
$^{1}$Institute for Astronomy, University of Edinburgh, Royal Observatory, Blackford Hill, Edinburgh, EH9 3HJ, UK\\
$^{2}$Centre for Extragalactic Astronomy, Department of Physics, Durham University, South Road, Durham DH1 3LE, UK\\
$^{3}$Armagh Observatory and Planetarium, College Hill, Armagh, BT61 9DG, N. Ireland, UK\\
$^{4}$Instituto de F\'isica y Astronom\'ia, Universidad de Valpara\'iso, Avda. Gran Breta\~na 1111, Valpara\'iso, Chile\\
$^{5}$Millennium Nucleus for Galaxies (MINGAL)\\
$^{6}$Institute of Science and Technology Austria (ISTA), Am Campus 1, 3400 Klosterneuburg, Austria\\
$^{7}$ School of Physics and Astronomy, University of Southampton, Highfield SO17 1BJ, UK\\
$^{8}$School of Physics and Astronomy, Lancaster University, Lancaster, LA1 4YB, UK
}

\date{Accepted XXX. Received YYY; in original form ZZZ}

\pubyear{\the\year{}}

\begin{document}
\label{firstpage}
\pagerange{\pageref{firstpage}--\pageref{lastpage}}
\maketitle


\begin{abstract}
The Ly$\alpha$ escape fraction, $f_{\mathrm{esc}}^{\mathrm{Ly}\alpha}$, probes both the interstellar medium (ISM) conditions governing ionizing photon escape and the rising neutral fraction of the IGM through the Epoch of Reionization (EoR). Characterising the intrinsic, ISM-driven distribution of $f_{\mathrm{esc}}^{\mathrm{Ly}\alpha}$ before IGM attenuation becomes dominant is essential to interpret the observed decline in Ly$\alpha$ visibility through the EoR. We present $f_{\mathrm{esc}}^{\mathrm{Ly}\alpha}$ measurements for a highly complete, H$\alpha$-flux-limited sample of 24 star-forming galaxies at $z \approx 6.1$, drawn from the \textit{JWST} Emission Line Survey (JELS) and observed in Ly$\alpha$ with VLT/MUSE as part of the JELS-MUSE Large Area Survey. We detect Ly$\alpha$ in 12 of 24 sources ($50 \pm 10$ per cent) and a Ly$\alpha$ emitter fraction of $X_{\mathrm{Ly}\alpha} = 33 \pm 12$ per cent using the canonical EW(Ly$\alpha$) $> 25$\,\AA\ definition. Incorporating non-detections via reverse Kaplan-Meier survival analysis yields $\langle f_{\mathrm{esc}}^{\mathrm{Ly}\alpha} \rangle = 0.07^{+0.04}_{-0.03}$, consistent with an independent stacked-flux estimate of $0.08^{+0.02}_{-0.02}$. Using reionization simulations matched to the area, depth, and redshift range of our survey, we find that all galaxies are expected to experience broadly similar IGM transmission, so we postulate that the large scatter in $f_{\mathrm{esc}}^{\mathrm{Ly}\alpha}$ reflects genuine ISM-driven variance rather than differences in the surrounding IGM. Among the detections, higher $f_{\mathrm{esc}}^{\mathrm{Ly}\alpha}$ galaxies tend to have lower nebular dust attenuation, bluer UV slopes, and lower stellar mass, consistent with feedback-regulated escape through localised, low-column-density ISM channels around star-forming regions. These results benchmark intrinsic Ly$\alpha$ escape at the end of reionization, against which IGM suppression at $z \gtrsim 7$ can be interpreted.
\end{abstract}



\section{Introduction}

The Epoch of Reionization (EoR) marks the last major phase transition of the Universe, during which hydrogen in the intergalactic medium (IGM) went from a predominantly neutral to a highly ionized state \citep[e.g.][]{Becker2001, Fan2006, Robertson2022}. Although star-forming galaxies are believed to be the primary sources responsible for driving reionization \citep[e.g.][]{Robertson2015, Finkelstein2019, Jiang2022}, precisely which populations dominate the process \citep{Naidu2020, Atek2024} and the exact contribution of active galactic nuclei (AGN) remain poorly understood, with some studies predicting a substantial or even dominant AGN contribution \citep[e.g.][]{Stevans2014,Madau2015,Smith2024,Cohon2026}. Fully resolving which sources are responsible requires a comprehensive understanding of their contributions to the ionizing photon budget, which depends on three key quantities \citep[e.g.][]{Robertson2013, Bouwens2016}: the ionizing photon production efficiency ($\xi_\mathrm{ion}$), the UV luminosity density ($\rho_\mathrm{UV}$), and the fraction of Lyman continuum photons (LyC; $\mathrm{h\nu\geq13.6\,eV}$) that escape into the IGM ($f_\mathrm{esc}^\mathrm{LyC}$). Together, these determine the rate at which ionizing photons are injected into the IGM. 

Linking these quantities to the physical properties of reionization-era galaxies is therefore key to identifying which sources drove the transition. Of these, $f_\mathrm{esc}^\mathrm{LyC}$ remains the least well constrained, as direct LyC observations become very difficult at $z \gtrsim 4.5$ due to the increasing opacity of the IGM \citep[e.g.][]{Inoue2014, McCandliss2017, Steidel2018,Goovaerts2026}. Even at intermediate redshifts ($z\sim2-3$), direct LyC measurements remain challenging due to stochastic attenuation by residual neutral hydrogen in the IGM and potential contamination from lower-redshift interlopers along the line-of-sight \citep[e.g.][]{Inoue2014, Steidel2018, Pahl2021, Begley2022}. 

In the absence of direct LyC measurements at high redshift, indirect tracers are essential.  Direct constraints on LyC escape have been established through low-redshift ($z\approx0.2-0.45$) LyC-emitting galaxies \citep[e.g.][]{Izotov2016a, Izotov2016, Izotov2018, Izotov2018a, Wang2019, Izotov2021}, and more systematically through the Low-redshift Lyman Continuum Survey \citep[LzLCS+;][]{Flury2022a, Flury2022b, Flury2025,Chisholm2022,Jaskot2024,Saldana-Lopez2022, Jaskot2025}. LzLCS+ was designed to target galaxies with high [\ion{O}{iii}]$\lambda5007$/[\ion{O}{ii}]$\lambda3727$ (O32) ratios and large H$\beta$ equivalent widths (EW) and has since established empirical correlations between LyC escape and a wider set of observable galaxy properties, many of which were already suggested by studies at $z \sim 3$ \citep[e.g.][]{Reddy2016,Steidel2018}. These include low neutral-gas covering fractions, blue UV slopes, low stellar masses and metallicities, compact star formation, and strong Lyman-alpha (Ly$\alpha$) emission and escape \citep[e.g.][]{Verhamme2017, Kimm2019, Flury2022b, Maji2022, Saldana-Lopez2026}. Among these diagnostics, Ly$\alpha$ has emerged as one of the strongest tracers of LyC escape. Because Ly$\alpha$ photons scatter resonantly in neutral hydrogen, their escape is regulated by the same neutral gas column density, covering fraction, and dust geometry that govern LyC escape \citep[e.g.][]{Verhamme2006,  Dijkstra2014, Verhamme2015, Gazagnes2020}. Observational studies have demonstrated a positive correlation between the escape of Ly$\alpha$ and LyC photons \citep[e.g.][]{Verhamme2015, Dijkstra2016, Izotov2020,Flury2022b, Begley2024, Choustikov2024}, making the Ly$\alpha$ escape fraction ($f_\mathrm{esc}^\mathrm{Ly\alpha}$) one of the most accessible probes of ionizing photon escape at high redshift \citep[e.g.][]{Marchi2018, Pahl2021, Pahl2024}.  Beyond the ISM, Ly$\alpha$ is also sensitive to the neutral hydrogen content of the IGM itself, since scattering along the line-of-sight means the ionization state of the surrounding IGM affects the observed Ly$\alpha$ \citep{Hayes2011, Matthee2022, Jaskot2025}. 

The visibility of Ly$\alpha$ emission is therefore shaped by two distinct physical regimes. At $z \lesssim 6$, the IGM is highly ionized, such that the observed $f_\mathrm{esc}^\mathrm{Ly\alpha}$ is governed primarily by ISM conditions within the galaxy itself, although attenuation from the Ly$\alpha$ forest remains non-negligible along the line-of-sight. In this regime, studies have demonstrated a steady increase in $f_\mathrm{esc}^\mathrm{Ly\alpha}$ with redshift \citep[e.g.][]{Hayes2011, Konno2016, Goovaerts2024}, often attributed to evolving ISM conditions across this range toward lower dust content, higher ionization parameters, and more porous gas geometries. However, this population-averaged trend need not reflect a smooth evolution within individual galaxies. \citet{Begley2024} find a negligible redshift evolution in $f_\mathrm{esc}^\mathrm{Ly\alpha}$ for individual galaxies once intrinsic scatter is accounted for, with a stronger dependence instead on UV magnitude, suggesting that much of the observed rise may be driven by a changing mix of galaxy properties within the population rather than evolution in the ISM conditions of any given galaxy. Consistent with this, samples of Ly$\alpha$ emitters selected with fixed criteria show approximately constant $f_\mathrm{esc}^\mathrm{Ly\alpha}$ \citep[e.g.][]{Matthee2022, Chen2024, Shimizu2025}, indicating that the global rise could instead be driven by an increasing fraction of galaxies with efficient Ly$\alpha$ escape at higher redshift. Regardless of the underlying cause, the population-averaged $f_\mathrm{esc}^\mathrm{Ly\alpha}$ rises towards higher redshift.

At $z \gtrsim 6.5$--$7$, however, the IGM becomes sufficiently neutral that Ly$\alpha$ transmission begins to drop \citep{Saxena2024}, and the measured $f_\mathrm{esc}^\mathrm{Ly\alpha}$ is increasingly dominated by IGM attenuation rather than intrinsic galaxy properties \citep[e.g.][]{Laursen2011, Mason2018, Gronke2021}. Recent observations have detected Ly$\alpha$ at $z \gtrsim 7$, where emitters are thought to reside in ionized bubbles in an otherwise neutral IGM \citep[e.g.][]{Saxena2024, Tang2024, Chen2024,Napolitano2024, Witstok2024, Witstok2026}, but disentangling the IGM and ISM contributions to the observed Ly$\alpha$ flux in this regime is highly uncertain \citep{Hayes2023}. Characterising the distribution and scatter of $f_\mathrm{esc}^\mathrm{Ly\alpha}$, and its dependence on physical galaxy properties at $z\sim6$ in the transition between the ISM-dominated and IGM-dominated regimes therefore provides a critical benchmark at the conclusion of the EoR against which the IGM-attenuated measurements at higher redshift can be interpreted.

Robustly measuring $f_\mathrm{esc}^\mathrm{Ly\alpha}$ requires an estimate of the intrinsic Ly$\alpha$ flux, most reliably inferred from H$\alpha$ emission under the standard assumption of Case B recombination. Prior to \textit{JWST}, H$\alpha$ from galaxies at $z \gtrsim 5$ was accessible only through Spitzer/IRAC broadband colour excesses, a method subject to substantial degeneracies and photometric uncertainties \citep[e.g.][]{Faisst2019, Stefanon2022}. \textit{JWST} has fundamentally changed this, with slitless spectroscopy and narrow-band imaging now providing improved H$\alpha$ flux measurements for large samples at these redshifts \citep[e.g.][]{Ning2023, Simmonds2023}, although the heterogeneous selection functions of current surveys can complicate completeness assessments. Deep NIRSpec spectroscopy has also enabled direct H$\alpha$ detections in individual galaxies \citep[e.g.][]{Saxena2023, Napolitano2024}. Integral field spectroscopy from the Very Large Telescope (VLT) Multi Unit Spectroscopic Explorer (MUSE; \citealt{Bacon2010}) complements these rest-optical observations well, offering continuous optical coverage from $4750$--$9350$\,\AA\ and no slit-loss effects, which enables sensitive Ly$\alpha$ observations across $2.9 < z < 6.7$ \citep[e.g.][]{Bacon2017, Leclercq2017, Kusakabe2022, Bacon2023}. \citet{Lin2024} demonstrated the power of combining these two facilities, conducting for the first time a direct census of $f_\mathrm{esc}^\mathrm{Ly\alpha}$ for an H$\alpha$-flux-limited sample at $z \approx 4.9$--$6.3$. \citet{Shimizu2026} also utilised  the benefits of a direct H$\alpha$-anchored approach by obtaining an analogous direct census at $z \simeq 6.2$ using dual narrow-band imaging, pairing \textit{JWST}/NIRCam F470N with Subaru/HSC NB872 to simultaneously select H$\alpha$ emitters and measure their Ly$\alpha$ flux. Most recently, \citet{Cheng2026} obtained a comparable $\mathrm{H}\alpha$-anchored census at $z \simeq 5.5$, pairing a \textit{JWST}/NIRCam F430M medium-band excess selection with archival VLT/MUSE $\mathrm{Ly}\alpha$ spectroscopy in the lensed A2744 field.

The key advantage of this H$\alpha$-selected approach is improved completeness. Ly$\alpha$-selected samples are directly biased towards galaxies with already favourable conditions for Ly$\alpha$ escape, missing galaxies with weak or undetected Ly$\alpha$ emission entirely. Lyman-break-selected samples avoid this particular bias, since selection is based on the UV continuum rather than Ly$\alpha$ strength directly, but such samples are typically UV flux limited, and Ly$\alpha$ equivalent width has been shown to anti-correlate with UV luminosity \citep{Stark2010}, so this selection can instead bias against the galaxies with the strongest Ly$\alpha$ escape. An H$\alpha$-selected sample avoids both of these effects and, although it still imposes a star-formation-rate floor on the sample given the shared origin of Ly$\alpha$ and H$\alpha$ in recent star formation, it provides the most complete census of $f_\mathrm{esc}^\mathrm{Ly\alpha}$ achievable across the star-forming galaxy population \citep{Oteo2015, Lin2024}. However, such studies remain rare, and existing measurements of $f_\mathrm{esc}^\mathrm{Ly\alpha}$ at $z \approx 5.5$--$6.5$ are largely derived from Ly$\alpha$, UV, or photometrically selected galaxies spanning different fields and redshifts, making it difficult to isolate intrinsic ISM drivers of Ly$\alpha$ escape from sample-selection effects or field-to-field  variations in IGM transmission. Furthermore, population-averaged estimates often rely on integration over Ly$\alpha$ or UV luminosity-functions, introducing systematic uncertainties from assumptions about the faint-end slope and incompleteness corrections \citep[e.g.][]{Hayes2011, Konno2016, Goovaerts2024}.  As a result, it remains unclear what drives the scatter in $f_\mathrm{esc}^\mathrm{Ly\alpha}$ at this epoch, and how much of the observed variation reflects intrinsic ISM conditions rather than IGM transmission. Current evidence suggests that H\,\textsc{i} column density, dust content, gas geometry, and gas kinematics (e.g. outflows and turbulence) all likely contribute to regulating Ly$\alpha$ escape, with correlations between Ly$\alpha$ escape and Ly$\alpha$ velocity separation indicating a particularly important role for line-of-sight neutral gas column density \citep{Verhamme2015, Gazagnes2020}. Variations in the ionizing photon production efficiency, $\xi_\mathrm{ion}$, and stellar population metallicity may also contribute to the observed scatter \citep{Sobral2019, Begley2026}. With observations of Ly$\alpha$ at $z \gtrsim 7$ now becoming more prevalent \citep[e.g.][]{Tang2023, Saxena2024,Jung2024, Jones2025,  Heintz2025,Napolitano2026}, a robust empirical benchmark for intrinsic Ly$\alpha$ escape near the end of the EoR is increasingly needed to interpret these higher-redshift measurements.

In this work, we address these challenges by measuring $f_\mathrm{esc}^\mathrm{Ly\alpha}$ for a highly complete, H$\alpha$-selected sample of star-forming galaxies at ${z \approx 6.1}$, combining \textit{JWST} narrow-band H$\alpha$ imaging with VLT/MUSE integral-field Ly$\alpha$ spectroscopy within a single contiguous field. Our H$\alpha$-selected parent sample is drawn from the \textit{JWST} Emission Line Survey \citep[JELS;][]{Duncan2025, Pirie2025}, an untargeted narrow-band NIRCam programme targeting H$\alpha$ emitters at $z \approx 6.05$--$6.25$ over a contiguous 63~arcmin$^2$ field in COSMOS-CANDELS, with no prior selection on Ly$\alpha$ properties. Ly$\alpha$ spectroscopy is provided by the JELS-MUSE Large Area Survey (Li et al., in prep.), one of the largest contiguous MUSE surveys to date, targeting a 63~arcmin$^2$ region overlapping the JELS footprint and delivering integral-field Ly$\alpha$ coverage for all JELS-selected sources within the survey area. Together, these datasets provide a uniquely complete sample drawn from a single large-scale environment. 


This paper is structured as follows. Section~\ref{sec:data} describes the JELS and MUSE datasets and the construction of our H$\alpha$-selected sample. Section~\ref{sec:COSMOSLAS} details the Ly$\alpha$ extraction, continuum subtraction, and source detection procedures, including the false-positive analysis used to define our detection threshold, and the measurement of Ly$\alpha$ fluxes, equivalent widths, and escape fractions. Section~\ref{sec:results} presents the Ly$\alpha$ detection fraction of our H$\alpha$-selected sample and examines what distinguishes the Ly$\alpha$-detected galaxies from the non-detections, comparing the two populations across their global properties. Section~\ref{sec:lyaesc} then turns to the escape fraction itself. We derive the average $f_\mathrm{esc}^\mathrm{Ly\alpha}$ of the sample, use a reionization simulation to test the uniformity of IGM transmission across the field, place these results in the context of the redshift evolution of Ly$\alpha$ escape, and in Section~\ref{sec:proprelations} examine how $f_\mathrm{esc}^\mathrm{Ly\alpha}$ depends on individual galaxy properties. Section~\ref{sec:summary} summarises our conclusions. Throughout, we assume a flat $\Lambda$CDM cosmology with $H_0 = 70$~km~s$^{-1}$~Mpc$^{-1}$, $\Omega_\mathrm{m} = 0.3$, and $\Omega_\Lambda = 0.7$. All magnitudes are quoted in the AB system \citep{Oke1983} and equivalent widths are quoted in the rest frame.

\section{H{\fontsize{14}{14}\selectfont$\mathbf{\alpha}$} observations and sample selection}
\label{sec:data}

\begin{figure}
	\includegraphics[width=\columnwidth]{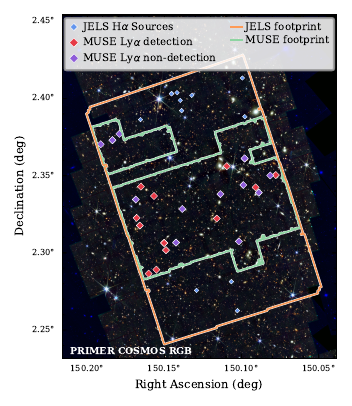}
	\caption{The COSMOS field coverage of the JELS-MUSE sample. The background shows the PRIMER NIRCam RGB composite. The three channels combine multiple filters, with blue from F090W, F115W and twice-weighted \textit{HST}/ACS F606W, green from F150W, F200W and F277W, and red from F356W and F444W. The MUSE IFU coverage available to be used in this study is outlined in green and the JELS F466N narrow-band mosaic in orange. Sources in the parent JELS H$\alpha$-selected sample are shown as small blue diamonds, with the subset observed with MUSE colour-coded by Ly$\alpha$ detection status, red for a significant Ly$\alpha$ detection and purple for a non-detection. The significance of a detection is defined in Section~\ref{sec:lyadetectsig}.}
	\label{fig:field_overview}
\end{figure}

\subsection{\textit{JWST} Emission Line Survey (JELS) photometry, catalogues and photometric redshifts}
\label{sec:jels}

The parent sample for this analysis is drawn from the H$\alpha$-emitter sample described in \citet{Pirie2025}. This sample utilises the \textit{JWST} Emission Line Survey \citep[JELS;][]{Duncan2025}, a narrow-band imaging programme conducted with \textit{JWST}/NIRCam. The survey covers a contiguous 63\,arcmin$^{2}$ centred at (RA, Dec) = (150.125$^\circ$, 2.333$^\circ$) and uses the F466N and F470N narrow-band filters at $\sim$4.7\,$\mu$m to identify H$\alpha$ emitters at $z \sim 6.1$. Figure~\ref{fig:field_overview} shows the outline of JELS with the JELS-MUSE Large Area Survey overlaid. Full details of the survey design, observing strategy and image reduction are given in \citet{Duncan2025} and so here we summarise only the aspects directly relevant to this analysis. The transmission curves for the F466N and F470N filters are shown in Figure~\ref{fig:filter_hist}, together with the H$\alpha$ redshift ranges they cover. In this work we use v1.0 of the JELS imaging and photometric catalogues, which incorporate re-observation of frames affected by scattered light along with a number of pipeline improvements, reaching 5$\sigma$ depths of 26.51 and 26.60 mag (measured in 0.3 arcsec diameter apertures) for the F466N and F470N images, respectively \citep[see Appendix A in][]{Duncan2025}.

A full description of the creation of the JELS narrow-band detected catalogues is provided in \citet{Pirie2025} and a brief summary is given here. Source detection was performed using \textsc{SExtractor} \citep{Bertin1996} in dual-image mode on the native-resolution narrow-band images, requiring the signal-to-noise ratio: $\mathrm{SNR} > 5$ within a 0.3 arcsec diameter aperture. Forced photometry was then measured on point-spread-function (PSF) homogenised images in multiple apertures (0.3 arcsec, 0.6 arcsec, 0.9 arcsec, 2.0 arcsec, and Kron; \citealt{Kron1980}). The 0.3 arcsec aperture photometry was used for colour-excess selection and photometric redshift analysis, while 0.6 arcsec measurements were adopted for line luminosity estimates and spectral energy distribution (SED) fitting \citep[][Section~\ref{sec:sed}]{Pirie2025}. To reduce contamination, \citet{Pirie2025} applied a minimum effective radius cut ($r_{\mathrm{e}} > 1.5$ pixels) to remove hot pixels and cosmic ray artefacts, and  also excluded sources located within bright-star masked regions.

Photometric redshifts (photo-$z$s) for all sources in the JELS photometric catalogues are derived using \texttt{EAZY-py} \citep{Brammer2008}, using all available \textit{JWST} and ancillary \textit{HST} photometry for a given source (see \citealt{Duncan2025} and \citealt{Pirie2025} for the full filter set and fitting procedure). Consensus redshift posteriors are derived following the Hierarchical Bayesian combination procedure of \citet{Duncan2018}, and the median of the primary 90 per cent highest probability density (HPD) credible interval is adopted as the redshift for each source. None of the sources in our final H$\alpha$ sample show a significant secondary photo-$z$ peak. Evaluated against spectroscopically confirmed sources, the photo-$z$ performance yields a normalised median absolute deviation of $\sigma_{\mathrm{NMAD}} = 0.030$ and an outlier fraction of 8.4 per cent across the full F466N/F470N detected source sample \citep{Pirie2025}.

\subsection{H{\fontsize{10}{10}\selectfont$\mathbf{\alpha}$} sample selection}
\label{sec:haselection}

H$\alpha$ emitters were identified following the selection criteria outlined by \citet{Pirie2025}, requiring a significant narrow-band excess relative to the F444W broadband imaging (BB$-$NB selection). All robust H$\alpha$ emitter candidates in this sample were recovered via this selection, with none requiring the alternative NB$-$NB approach. Excess significance was quantified using the NB excess parameter $\Sigma$, which was used to remove sources that likely have no intrinsic narrow-band colour excess, but which scatter to higher values due to local noise variations that become prominent for the faintest sources \citep[e.g. ][]{Bunker1995,Sobral2013}:

\begin{align}
    \Sigma = \frac{1 - 10^{-0.4(\mathrm{BB} - \mathrm{NB})}}{10^{-0.4(\mathrm{ZP} - \mathrm{NB})}\sqrt{\sigma_{\mathrm{NB}}^2 + \sigma_{\mathrm{BB}}^2}}
\end{align}

\noindent where ZP is the zero point magnitude of the NB filter, set to 23.9 mag, and $\sigma_{\mathrm{NB}}$ and $\sigma_{\mathrm{BB}}$ are the photometric flux density errors (in $\mu$Jy) for the NB and BB filters respectively for each source. Candidates are required to satisfy $\Sigma > 3$ and $\mathrm{F444W} - \mathrm{F466N} > 0.3$ for F466N detections, or $\Sigma > 3$ and $\mathrm{F444W} - \mathrm{F470N} > 0.3$ for F470N detections. The BB$-$NB colour cuts address further systematics, such as continuum colour corrections, which introduce a degree of scatter at bright magnitudes.

To isolate H$\alpha$ emission at $z \approx 6$, additional photo-$z$ constraints were imposed \citep{Pirie2025}:
\begin{align}
5.5 \leq z_{\mathrm{median}} \leq 6.5, \label{eq:zrange} \\
\frac{z_{\max} - z_{\min}}{1 + z_{\mathrm{median}}} < 0.4, \label{eq:zwidth}
\end{align}
where $z_{\rm min}$, $z_{\rm max}$, and $z_{\rm median}$ are the lower bound, upper bound, and median of the primary 90 per cent HPD credible interval respectively (Section~\ref{sec:jels}). These conditions are a robustness requirement, ensuring that the retained redshift posteriors are well constrained and narrow, as expected for genuine line-driven narrow-band excess. 
Implementing Equation~\ref{eq:zwidth} is required generally for robust selections of emission line galaxies, but it has no effect on the H$\alpha$ selection presented here.

\begin{figure}
	\includegraphics[width=\columnwidth]{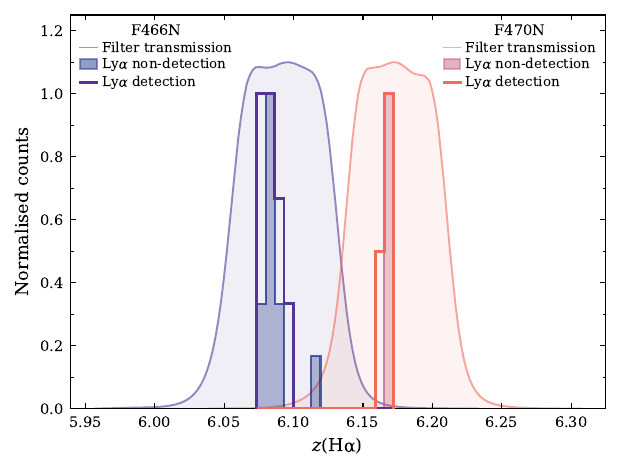}
	\caption{Redshift distribution of the JELS H$\alpha$-selected sample as a function of \textit{JWST}/NIRCam narrow-band filter. Histograms show sources targeted by F466N (purple) and F470N (red), split into Ly$\alpha$ detections (solid outline) and non-detections (shaded). Each histogram is scaled to unity to allow direct visual comparison. Overlaid curves show the F466N and F470N filter transmission profiles \citep[SVO Filter Profile Service:][]{Rodrigo2012, Rodrigo2020}, converted to an equivalent H$\alpha$ redshift scale. These curves are included for illustrative purposes to demonstrate where the source redshifts fall relative to the filter bandpasses and so are arbitrarily rescaled in the vertical direction.} 
	\label{fig:filter_hist}
\end{figure}

The updated selection using the improved v1.0 imaging reductions is presented in \citet{Pirie2026}, yielding 39 robust H$\alpha$ emitters from 44 candidates ($\sim$89 per cent purity). We use this updated sample throughout this work. Of these 39 sources, 27 fall within the current JELS-MUSE Large Area Survey footprint. These sources are marked on the RGB footprint in Figure~\ref{fig:field_overview}, and a histogram of the sources in their respective NIRCam filters is shown as a function of $z(\mathrm{H}\alpha)$ in Figure~\ref{fig:filter_hist}. Based on detailed completeness simulations that are driven primarily by the source detection SNR and narrow-band excess selection criteria, \citet{Pirie2026} demonstrate that the H$\alpha$ sample is $>80$ per cent complete down to
$\log(L_{\rm H\alpha}/{\rm erg\,s^{-1}}) \gtrsim 41.5$, and $>90$ per cent
complete above $\log(L_{\rm H\alpha}/{\rm erg\,s^{-1}}) \gtrsim 41.6$ (both
measured within a 0.3 arcsec diameter aperture).

Through detailed simulations of narrow-band selected samples, \citet{Duncan2025} demonstrate that the inclusion of the JELS F466N/F470N filters yields photo-$z$ scatter $\sigma_{\rm NMAD} \lesssim 0.005\times(1+z)$. The individual redshift posterior widths, spanning the 16th to 84th percentiles of the redshift probability distribution, are constrained to $\Delta z_{16-84} \lesssim 0.03$ for H$\alpha$ luminosities and equivalent widths (EW) typical of the JELS sample. This redshift uncertainty is significantly narrower than the effective redshift coverage of the F466N/F470N filters shown in Figure \ref{fig:filter_hist}, indicating that the photometric redshifts constrain sources to a relatively narrow region within each narrow-band selection window, rather than simply identifying them as lying somewhere within the filter passband. This represents an improvement of a factor of $2$-$5$ over equivalent medium-band or broadband-only estimates. This is also shown in \citet{Pirie2025}, where observed H$\alpha$ emitters at $z \approx 6.1$ have photo-$z$ uncertainties well within the narrow-band transmission profiles.

Direct support for this precision comes from the three sources in our sample with spectroscopic redshifts available through the Dawn \textit{JWST} Archive \citep[DJA;][]{deGraaff2024, Heintz2024}, for which the photometric and spectroscopic redshifts agree to within $\Delta z \lesssim 0.02$ (see Section~\ref{sec:lyadetectsig}). This precision motivates key steps in the Ly$\alpha$ selection described in Section~\ref{sec:lyadetectsig}, providing both the positional prior used to search for Ly$\alpha$ emission in the MUSE data and the luminosity distances used in all subsequent luminosity calculations. Figure~\ref{fig:filter_hist} shows that all sources have photometric redshifts placing H$\alpha$ close to the peak of the F466N or F470N transmission profiles, where the transmission is close to unity, so no filter transmission corrections are applied to the H$\alpha$ fluxes as corrections would have a negligible effect on these sources.

\subsection{H{\fontsize{10}{10}\selectfont$\mathbf{\alpha}$} Flux and Luminosity}
\label{sec:haflux}

H$\alpha$ line fluxes are calculated for each source in the JELS parent sample following \citet{Pirie2025} using the broadband-narrow-band (BB - NB) method \citep[e.g.][]{Sobral2013}:

\begin{equation}
    F_{\rm{H}\alpha} = \Delta\lambda_{\rm NB}\,
    \frac{f_{\rm NB} - f_{\rm BB}}
         {1 - \left(\Delta\lambda_{\rm NB}/\Delta\lambda_{\rm BB}\right)},
\end{equation}

\noindent where $\Delta\lambda_{\rm NB}$ and $\Delta\lambda_{\rm BB}$ are the widths of the narrow-band and broadband filters respectively, and $f_{\rm NB}$ and $f_{\rm BB}$ are the corresponding measured flux densities. Photometry measured in 0.6 arcsec diameter apertures is used, capturing $\sim$82 per cent of the total line flux for a point source while minimising contamination from neighbouring objects \citep{Pirie2025}. Aperture-to-total corrections are applied on a source-by-source basis by scaling to the Kron aperture photometry \citep{Kron1980}. The H$\alpha$ luminosity is then $L_{\rm{H}\alpha} = 4\pi D_{\rm L}^{2} F_{\rm{H}\alpha}$, where $D_{\rm L}$ is the luminosity distance evaluated at $z_{\rm median}$ for each source. The narrow-band flux includes a contribution from the [\ion{N}{ii}]$\lambda6583$ emission line. Following \citet{Shapley2023}, who measure $\log_{10}([\ion{N}{ii}]\lambda6583/\mathrm{H}\alpha) = -1.31$ with negligible [\ion{N}{ii}]$\lambda6548$ from spectral stacks of $z = 5.0$--$6.5$ star-forming galaxies, we apply a correction of $0.021$ dex to $L_{\mathrm{H}\alpha}$, consistent with the JELS H$\alpha$ measurements of \citet{Pirie2025} and \citet{Duncan2025}.

We identify two sources in the parent sample as probable AGN candidates on the basis of their high stellar masses inferred from SED fitting and compact, PSF-like morphologies \citep{Pirie2025, Stephenson2025}. These sources are retained, but explicitly flagged, in analyses of empirical observables, while they are excluded from analyses based on SED-derived physical properties, as the SED fitting (see Section \ref{sec:sed}) does not include AGN components and may therefore yield unreliable parameter estimates for these systems. We inspected the MUSE spectra of both sources for high-ionization emission lines. No significant \ion{N}{v}$\lambda$1240 is detected in either source. Adopting a \ion{N}{v}-to-Ly$\alpha$ flux ratio of 10 per cent \citep{VandenBerk2001}, the expected \ion{N}{v} signal at $\lambda_{\rm obs} \approx 8778$\,\AA\ corresponds to a predicted SNR of $<$2 in both cases, meaning a non-detection is fully consistent with the noise level even if these sources host AGN activity. The remaining common high-ionization diagnostics (\ion{C}{iv}\,$\lambda$1549, He\,\textsc{ii}\,$\lambda$1640, \ion{C}{iii}\,$\lambda$1909) fall beyond the MUSE wavelength range at $z \approx 6.1$ and are inaccessible.

\subsection{SED Fitting and Derived Galaxy Properties}
\label{sec:sed}

The derived galaxy properties used in this paper come from SED fitting performed for all sources in the JELS parent sample using the \textsc{BAGPIPES} code \citep{Carnall2018}, as described in \citet{Pirie2025}. A brief summary is given here. Stellar populations are modelled with the \textsc{BPASS v2.2} \citep{Eldridge2017,Stanway2018} stellar population synthesis (SPS) code, assuming a Kroupa-type initial mass function (IMF) \citep{Kroupa1993} with an upper mass cut-off of $100\,\mathrm{M_\odot}$, with nebular emission computed via the \textsc{Cloudy} photoionization code \citep{Ferland2017}. For the star-formation histories, we adopt the continuity non-parametric framework of \citet{Leja2019}, with bin edges at $[0, 3, 10, 30, 100, 300, 750]$\,Myr. Dust attenuation is modelled following \citet{Salim2018}, which parametrises deviations from the \citet{Calzetti2000} attenuation law via a slope offset $\delta$. The redshift prior for each source is set as a uniform distribution spanning the 16th--84th percentile range of its \texttt{EAZY-py} posterior (Section~\ref{sec:jels}), or $z_{\rm median} \pm 0.1$ where this range is narrower than $\pm0.1$. The fit uses the same multiwavelength \textit{JWST} and ancillary \textit{HST} photometry as the photometric redshift analysis.

From the SED fits we extract posterior distributions for the stellar mass $M_{\star}$, star-formation rates (SFR) averaged over 10\,Myr timescale ($\mathrm{SFR_{10}}$), and the $V$-band stellar continuum attenuation $A_{V}$, which is fitted directly as a free parameter in the SED fit rather than derived afterwards. The corresponding colour excess is computed as $E(B-V) = A_V/4.05$, following the \citet{Calzetti2000} attenuation law,  and under the assumption that reddening from the stellar continuum and nebular emission is the same (see discussion below), this is taken to represent both the stellar and nebular reddening. The SED fits are performed on $0.6$\,arcsec aperture photometry, so we apply aperture corrections to $M_\star$ and the SFRs using the ratio of Kron to $0.6$\,arcsec narrow-band flux for each source. Finally, specific star-formation rate ($\mathrm{sSFR_{10}}$) used throughout this work is defined as $\mathrm{SFR_{10}}$ divided by $M_\star$.

To dust-correct the H$\alpha$ luminosities, we scale the SED-derived $A_V$ by the \citet{Calzetti2000} attenuation curve to the H$\alpha$ wavelength of 6563\,\AA. \citet{Pirie2025} found negligible deviations from the \citet{Calzetti2000} attenuation slope on average for the H$\alpha$ sample (median $\delta = -0.1$), supporting this choice. \citet{Pirie2026} tested the impact of the assumed reddening factor between the stellar continuum and nebular emission, $\eta_{\mathrm{dust}} = A_{\mathrm{cont}}(6563\,\text{\AA})/A_{\mathrm{H}\alpha}$, on the SED-derived $A_V$ values, adopting $\eta_{\mathrm{dust}}=1$ (equivalent reddening) and $\eta_{\mathrm{dust}}=0.44$. This corresponds to the empirical continuum-to-nebular colour excess ratio, $E(B-V)_{\mathrm{cont}}/E(B-V)_{\mathrm{neb}}$, measured by \citet{Calzetti2000}, under the simplifying assumption that the same dust law applies to both the stellar continuum and nebular emission. Because $A_V$ is a free parameter in the fit, changing $\eta_{\mathrm{dust}}$ shifts the fitted stellar-continuum attenuation itself. Assuming $\eta_{\mathrm{dust}}=0.44$ results in a decrease in the SED-derived $A_V$ values compared to $\eta_{\mathrm{dust}}=1$, such that the resulting extinction on the H$\alpha$ emission line, $A_{\mathrm{H}\alpha}$, is similar between the two assumptions. Given the above discussion, we adopt $\eta_{\mathrm{dust}}=1$ throughout and refer the reader to \citet{Pirie2026} for a full discussion. 

The absolute UV magnitude $M_{\rm UV}$ and UV-continuum slope $\beta$ are measured empirically from the aperture-corrected photometry following \citet{Pirie2025}. The flux densities in each filter probing the rest-frame UV continuum ($\lambda_{\rm rest} \leq 3000$\,\AA, avoiding the Ly$\alpha$ line and IGM absorption) are fitted with a power law $f_{\lambda} = f_{0}(\lambda/\lambda_{0})^{\beta}$ using a non-linear least-squares approach weighted by photometric uncertainties, where $\lambda_{0} = 1500$\,\AA. The flux density at rest-frame 1500\,\AA\ is extracted from the fitted spectrum to compute $L_{1500}$ and hence $M_{\rm UV}$ \citep{Oke1983}. $M_{\rm UV}$ values used throughout this work are observed (no dust-corrections applied), consistent with standard practice in Ly$\alpha$ fraction studies \citep[e.g.][]{Stark2011,Mason2018}. 

\section{JELS-MUSE Ly{\fontsize{14}{14}\selectfont$\mathbf{\alpha}$} observations}
\label{sec:COSMOSLAS}

\begin{figure*}
    \includegraphics[width=\textwidth]{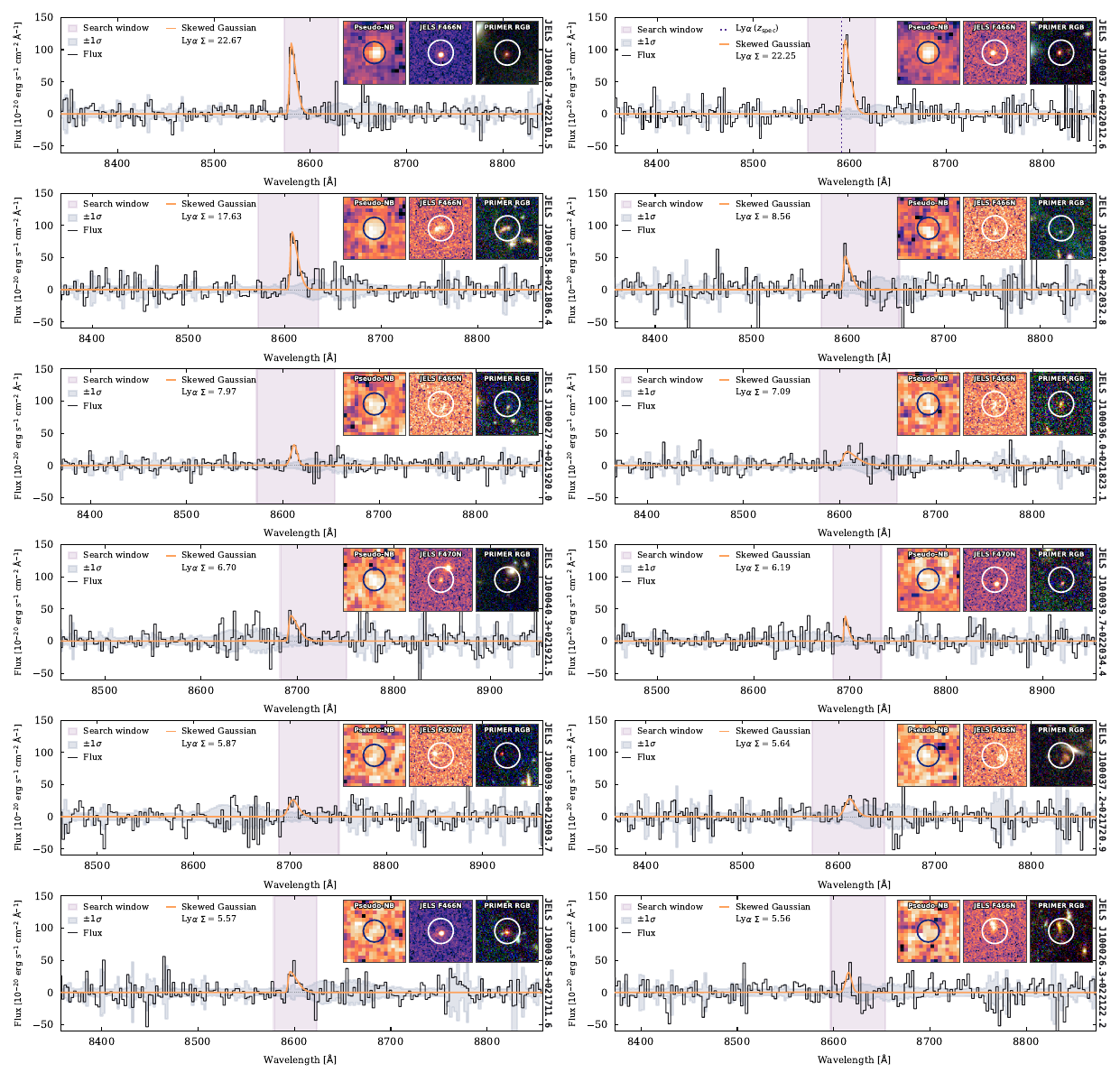}
    \caption{Ly$\alpha$ emission-line spectra of the Ly$\alpha$ detections ($\Sigma_{\mathrm{Ly}\alpha} > \Sigma_{\mathrm{Ly}\alpha}^{98} = 5.09$) from MUSE observations of the JELS field. Sources are ordered by descending Ly$\alpha$ significance, from top left to bottom right. The first seven sources exceed the strong-detection threshold ($\Sigma_{\mathrm{Ly}\alpha} > \Sigma_{\mathrm{Ly}\alpha}^{99.5} = 6.41$). Each panel shows the extracted flux spectrum (black step histogram) within a 0.6 arcsec circular aperture, with the $\pm 1\sigma$ noise envelope shown in grey. The shaded purple region indicates the photometric redshift search window derived from the JELS H$\alpha$ redshift constraints in Section~\ref{sec:lyadetectsig}. Sources with spectroscopic redshifts have this marked by a purple dotted vertical line and the search window fixed to this central wavelength. The best-fitting skewed Gaussian profile is overlaid in orange. Inset images ($3\ \mathrm{arcsec} \times 3\ \mathrm{arcsec}$) show, from left to right: a pseudo-narrow-band image collapsed over the Ly$\alpha$ detection window from the MUSE datacube; the JELS photometric detection in the relevant NIRCam band; and a PRIMER RGB composite (channel definitions as in Fig.~\ref{fig:field_overview}). A 0.6 arcsec aperture circle is overlaid on each cutout centred by the Ly$\alpha$ extraction point. We note that the aperture positions are set by optimising the Ly$\alpha$ significance across the spatial and spectral dimensions (Section~\ref{sec:lyadetectsig}), so a source may appear off centre in the two-dimensional narrow-band cutout but still be the most optimal extraction point. Source identifiers are given to the right of each spectrum. }
    \label{fig:detect}
\end{figure*}

The Multi Unit Spectroscopic Explorer (MUSE) is an optical integral-field spectrograph mounted on the Very Large Telescope (VLT). In Wide Field Mode, MUSE provides a 1\,arcmin$^{2}$ field of view with 0.2  arcsec spatial sampling and continuous spectral coverage from 4750 to 9350\,\AA\ at a mean resolving power of $R \sim 3000$ ($R = 1770$ at 480 nm to $R = 3590$ at 930 nm), and sampled at 1.25 \AA\, per spectral pixel in the output cube. This wavelength range enables the detection of Ly$\alpha$ emission from galaxies over $2.9 < z < 6.7$, encompassing the end stages of the EoR that this study focuses on. 

The JELS-MUSE Large Area Survey (PID: 112.25WM) is one of the largest contiguous MUSE surveys to date. The survey targets a contiguous 63\,arcmin$^{2}$ region within the COSMOS CANDELS field, selected to overlap the \textit{JWST} PRIMER and JELS imaging footprints. The final mosaic will consist of 77 individual MUSE pointings. At the time of this analysis, 34 pointings (44 per cent) have been completed and fully reduced, with a further 21 observed in 2026. Observations were prioritised such that each pointing is completed before moving on to the next pointing and so all the observations used in this analysis are at their full depth. The exposure depth per pointing was designed to enable the detection of Ly$\alpha$ emission from $z \approx 6.1$ H$\alpha$-selected galaxies identified by JELS, ensuring sensitive and spatially resolved spectroscopy across the full survey footprint. The completed and reduced datacubes used in this analysis are highlighted in Figure~\ref{fig:field_overview}, along with the JELS H$\alpha$ sources both inside and outside the MUSE coverage. The observation schedule of the MUSE survey was set to prioritise pointings with known JELS H$\alpha$ sources, so we have a higher fraction of the parent H$\alpha$ sample within the current footprint than would be expected from a representative subset of pointings. A full description of the survey design and data products will be presented in Li et al.\ (in prep.).

A brief summary of the observations and data reduction procedure is provided here. Each MUSE pointing comprises three observing blocks (OBs), with each OB split into three 900 second exposures, reaching a full depth of 2.25h on-source time per pointing. The three exposures within each OB are offset in RA/Dec by $\sim 8''$ and rotated through 0, 90, and 180$^\circ$, to account for flat-field variations both along and between the IFU slices. We use the wide-field adaptive optics (AO) mode, which typically improves the PSF full width at half maximum (FWHM) by a factor $\sim 2$ compared to natural seeing, although five pointings lack suitable AO guide stars and were instead executed in seeing-limited (no AO) mode. All observations were taken in clear or photometric conditions with natural $V$-band seeing $< 0.7''$, between December 2023 and May 2025. 

The data were reduced with the MUSE pipeline \citep{Weilbacher2020}, run within the \textsc{EsoReflex} workflow environment version 3.13.8 \citep{Freudling2013}. The pipeline performs standard bias, dark current, flat-field, and sky-flat corrections, along with wavelength calibration using arc lamp exposures, flux calibration using nightly standard stars, and geometric and astrometric calibration, and it corrects for atmospheric and telluric extinction. For each science frame, additional flat fielding is performed using the \textsc{esorex} auto-calibration mode, which corrects residual inter-slice illumination offsets and refines the sky subtraction across all the IFU slices. To further improve the sky subtraction, we process the autocalibrated exposures with the Zurich Atmosphere Purge \citep[ZAP;][]{Soto2016}. For each exposure we generate a source mask by collapsing the datacube over its full wavelength range and running \textsc{SExtractor} \citep{Bertin1996} to identify regions free from stars and galaxies in the resulting white-light image. This mask is passed to ZAP to exclude continuum sources from the sky model, and a principal-component analysis is used to model and subtract the residual sky emission in every pixel. Appendix~\ref{appendix:astrometry} details the astrometric corrections applied to the reduced cubes and the mosaicking technique used to construct an interim combined data product of the full field.

 Of the 27 H$\alpha$ emitters in the JELS-MUSE footprint, three pairs of sources are spatially unresolved by MUSE and are therefore merged, yielding a parent catalogue of 24 sources for our subsequent H$\alpha$--Ly$\alpha$ analysis. For merged pairs, integrated properties are taken as the sum of the individual measurements, while specific or relative properties are computed as the H$\alpha$ luminosity-weighted mean of the two components. All 24 sources are marked on the RGB footprint in Figure~\ref{fig:field_overview} and also shown on the redshift histogram in Figure~\ref{fig:filter_hist}, with corresponding positions within the transmission profile shown for the relevant detection filter. MUSE spectra with accompanying pseudo-NB, JELS NB, and PRIMER RGB cutouts are shown for the detections in Figure~\ref{fig:detect} and non-detections are shown in Figure~\ref{fig:nondetect}.

\subsection{Ly{\fontsize{10}{10}\selectfont$\mathbf{\alpha}$} Source Extraction}
\label{sec:extraction}

We search for Ly$\alpha$ emission from all JELS H$\alpha$-selected galaxies within the MUSE footprint using a targeted approach that exploits the known spatial positions and photometric redshifts of the sources. For each galaxy, we extract a $20\ \mathrm{arcsec} \times 20\ \mathrm{arcsec}$ subcube centred on the \textit{JWST} position from the fully reduced MUSE mosaic (see Appendix \ref{appendix:astrometry}). Each subcube spans the full spectral range of the data (4749.9--9349.9\,\AA, sampled at $\Delta\lambda = 1.25$\,\AA).

The expected Ly$\alpha$ wavelength was computed from the median photometric redshift. Continuum emission was removed using a spectral median-filtering approach applied independently to each spaxel (a single spatial pixel of the cube, with its own full spectrum), following methods commonly adopted in deep MUSE analyses \citep[e.g.][]{Herenz2017,Leclercq2017, Kusakabe2022}. For each spaxel, spectral channels (individual wavelength slices of the cube) within the photometric redshift range of the source (from $z_{\mathrm{min}}$ to $z_{\mathrm{max}}$ with an additional 5\,\AA\ buffer on either side) were excluded from the continuum estimate. This window is wider than the fixed $\pm 400$\,km\,s$^{-1}$ mask adopted by studies with spectroscopic redshifts \citep[e.g.][]{Kusakabe2022} and accounts for the photometric redshift uncertainty on the Ly$\alpha$ position, ensuring line flux does not contaminate the continuum estimate.

The remaining spectral channels were used to construct a continuum model via a one-dimensional median filter with a fixed window size of 101 spectral pixels ($\sim$126\,\AA). This window is much broader than the width of any targeted emission line, so genuine line flux is not removed by the filter, while remaining narrow enough to track the slowly-varying stellar continuum of each source. To ensure stability of the filtering in the presence of masked channels and non-finite values, excluded pixels were temporarily filled with the median flux of the valid continuum region prior to filtering. After median filtering, the continuum model was linearly interpolated across masked wavelength regions, allowing smooth extrapolation through the Ly$\alpha$ interval.

Continuum subtraction was applied to each spatial pixel (spaxel) in the subcubes to generate a continuum-subtracted cube for each source. Because the filter window is much wider than an emission line but narrow enough to follow the stellar continuum shape, it isolates and removes this continuum while leaving line emission intact. The procedure was applied uniformly across all spaxels, enabling unbiased searches for Ly$\alpha$ emission both at the source position and in surrounding regions for background and false-positive characterisation. Visual inspection confirmed that the subtraction does not introduce systematic residuals at the expected Ly$\alpha$ wavelength.

\subsection{Determining Optimal Ly{\fontsize{10}{10}\selectfont$\mathbf{\alpha}$} Measurements}
\label{sec:cog}

To derive Ly$\alpha$ escape fractions, robust Ly$\alpha$ flux measurements are required for each source. We therefore first determine an optimal spatial aperture for spectral extraction based on both physically motivated priors and empirical tests. 

Previous studies of spatially extended Ly$\alpha$ emission show that growth curves typically continue to rise to relatively large radii (often $r \gtrsim 3\ \mathrm{arcsec}$), reflecting the presence of low-surface-brightness emission beyond the core \citep[e.g.][]{Wisotzki2016}. However, because our goal is to measure Ly$\alpha$ escape fractions in the context of reionization, capturing the full spatial extent of the halo is not necessarily optimal. Extended Ly$\alpha$ emission is known to arise from multiple physical processes beyond \textit{in situ} recombination within the central galaxy. Observational studies have shown that Ly$\alpha$ haloes may include contributions from cooling radiation, Ly$\alpha$ fluorescence powered by the ultraviolet background, and emission from nearby satellite galaxies \citep[e.g.][]{Leclercq2017,Kusakabe2020,HerreroAlonso2023}. While it is difficult to observationally disentangle these components, simulations suggest that within the inner $\lesssim$5-7 kpc at $3 < z < 6$, Ly$\alpha$ emission is dominated by scattering of recombination radiation produced within the galaxy and its immediate circumgalactic environment \citep[e.g.][]{Mitchell2021}. At larger radii, the relative importance of non-recombination processes and environmental contributions increases \citep{Tang2024a}. We note that recent work has found a strong anti-correlation between LyC escape fraction and the fraction of Ly$\alpha$ flux observed in extended haloes \citep[e.g.][]{Saldana-Lopez2026}, indicating that halo emission may itself contain information about LyC leakage. This anti-correlation may partly reflect a physical decoupling between the two escape channels on large scales: Ly$\alpha$ photons that reach the halo can still escape after repeated scattering, whereas LyC photons are removed by absorption rather than redirected, so extended Ly$\alpha$ emission need not imply a correspondingly extended reservoir of escaping ionizing photons.

\begin{figure}
    \includegraphics[width=\columnwidth]{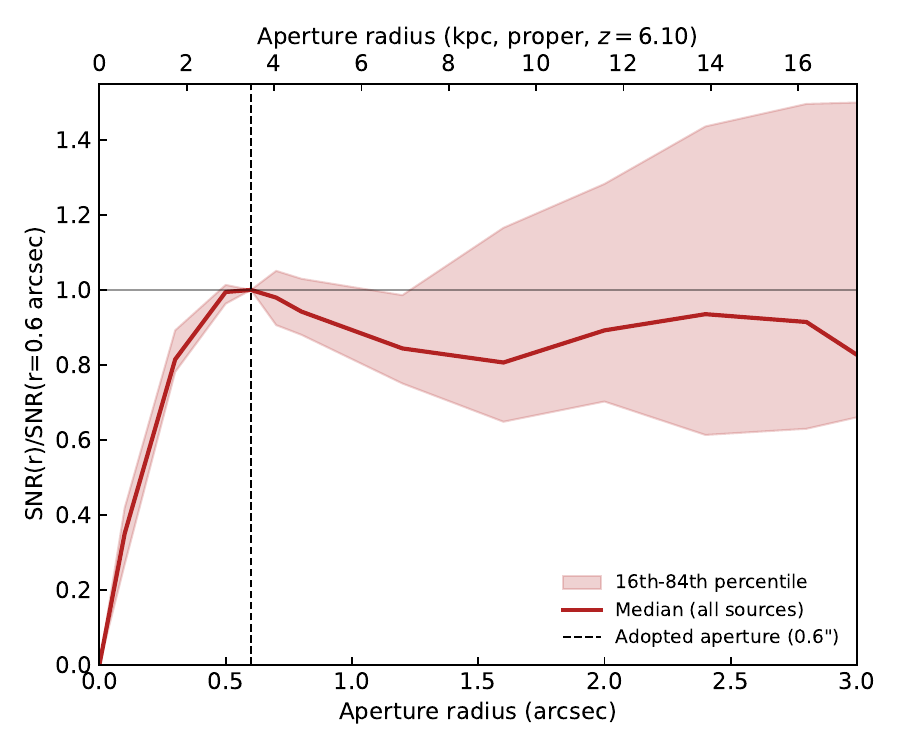}
    \caption{Stacked SNR curve of growth for the 12 Ly$\alpha$ detections. Each source's SNR is normalised to its own value at $r = 0.6$ arcsec, so all curves equal unity at the adopted aperture (dashed line) by construction. The solid line is the sample median and the shaded region the 16th to 84th percentile range. The upper axis gives the proper transverse scale at $z = 6.10$. The median remains at or below unity out to $r = 3$ arcsec, indicating no SNR gain from a larger aperture.}
    \label{fig:curveofgrowth}
\end{figure}

We therefore use curve-of-growth analysis to prioritise an aperture that isolates the central, recombination-dominated Ly$\alpha$ emission rather than the full low-surface-brightness halo. This involves measuring the enclosed Ly$\alpha$ flux and SNR as a function of circular aperture radius for each source, centred on their JELS-predicted Ly$\alpha$ positions. The SNR growth curves exhibit a clear local maximum at small radii. Across the sample, this inner peak occurs at $r \approx 0.6\ \mathrm{arcsec}$, beyond which the inclusion of additional halo emission leads to an initial decline in SNR. Figure~\ref{fig:curveofgrowth} shows this behaviour for the sample as a whole, with each curve normalised to its own value at $r = 0.6\ \mathrm{arcsec}$. The median curve rises steeply out to $r \simeq 0.6\ \mathrm{arcsec}$ and then remains at or below unity across the full range plotted, so no larger aperture recovers additional SNR for the typical source. Individual sources scatter about the median and the 16th to 84th percentile range broadens at large radii, where the enclosed noise grows and the emission becomes increasingly faint. At $z \approx 6$, a radius of 0.6 arcsec corresponds to $\approx$ 3-4 kpc (upper axis of Figure~\ref{fig:curveofgrowth}), well within the $\lesssim$5-7 kpc regime expected to be dominated by internally produced Ly$\alpha$ emission. Placing the aperture instead at the outer edge of that regime, $r \approx 1.0-1.2\ \mathrm{arcsec}$, would sit where the median SNR is lower by $\approx 15$ per cent while adding halo emission of increasingly uncertain origin. An aperture of $r = 0.6\ \mathrm{arcsec}$ still encompasses the inner Ly$\alpha$ halo (including the immediate circumgalactic medium, CGM), while reducing sensitivity to the more extended low-surface-brightness emission at larger radii. We therefore adopt a fixed circular aperture of $r = 0.6\ \mathrm{arcsec}$ for Ly$\alpha$ spectral extraction throughout the remainder of this work.

\subsection{Ly{\fontsize{10}{10}\selectfont$\mathbf{\alpha}$} Detection and Significance}
\label{sec:lyadetectsig}

\subsubsection{Spatial centroid optimisation}

Due to resonant scattering, Ly$\alpha$ emission can be spatially offset from the UV stellar continuum. Astrometric offsets at the level of a MUSE pixel (0.2 arcsec) are also possible (see Appendix~\ref{appendix:astrometry}). Spatial offsets between Ly$\alpha$ and rest-frame UV continuum emission are typically $0.2-2$ kpc across $2\lesssim z\lesssim6$ \citep[e.g.][]{Shibuya2014, Hoag2019, Ribeiro2020, Khusanova2020, Lemaux2021, Claeyssens2022, Ning2024}, with \citet{Claeyssens2022} reporting a median centroid offset of 0.58 kpc. There is less literature available on Ly$\alpha$-H$\alpha$ offsets, though the same resonant scattering responsible for Ly$\alpha$-UV offsets would plausibly produce a similar effect relative to H$\alpha$, since H$\alpha$ traces the same star-forming regions as the UV continuum. This makes an empirical, significance-maximising search for the true Ly$\alpha$ position a useful step regardless of which continuum tracer is used as the positional prior. We therefore perform a grid search in right ascension and declination around the H$\alpha$ prior position to maximise the statistical significance of the integrated flux for the extracted Ly$\alpha$ emission. For each source we search on a grid with step size 0.1 arcsec out to a maximum offset of 0.4 arcsec, corresponding to approximately twice the typical astrometric residual (Appendix~\ref{appendix:astrometry}), using a circular aperture of 0.6 arcsec radius. To prevent the optimisation from settling on noise fluctuations far from the expected location, the significance at each trial position is multiplied by a Gaussian weighting function, centred on the JELS position, with a fixed width of 1 arcsec. This weighting favours trial positions closer to the JELS prior whenever their significance is comparable to that of a more distant position. The prior and the 0.4 arcsec cap together keep the false-positive rate low, at the cost of limiting how far the aperture can move towards the Ly$\alpha$ peak. We also tested wider search radii and found that the set of significant detections is unchanged, indicating that the 0.4 arcsec cap is not excluding genuine Ly$\alpha$ signal.

Because the extraction position is chosen numerically in three dimensions, by maximising the integrated Ly$\alpha$ significance across the aperture and a localised spectral window, it is not necessarily centred on the UV continuum or on the brightest pixel in the collapsed cutout. Consequently, some sources appear visually offset in the two-dimensional cutouts shown in Figures~\ref{fig:detect} and~\ref{fig:nondetect}.  Allowing the aperture to move freely to the surface-brightness peak of each source does not measurably increase the recovered flux, and most sources instead show a small decrease due to source asymmetry. Since permitting larger offsets would raise the false-positive rate for no meaningful gain, we retain the significance-optimised positions within the 0.4 arcsec cap described above.

\subsubsection{Spectral window optimisation}

Once the optimal spatial centroid is determined, we search for the Ly$\alpha$ peak across the full predicted redshift range. Because Ly$\alpha$ emission can be offset in velocity from both the systemic redshift and the H$\alpha$-based redshift, the line centre may deviate from the photometric redshift prediction. We therefore slide a 10\,\AA\ spectral window in steps of one spectral pixel (1.25\,\AA) across the wavelength range shown in Figures~\ref{fig:detect} and~\ref{fig:nondetect}. This search window corresponds to the JELS photometric redshift bounds $[z_\mathrm{min}, z_\mathrm{max}]$, extended by 5\,\AA\ on either side to accommodate additional velocity offsets. We adopt the JELS photometric redshift bounds, rather than the wider range permitted by the H$\alpha$ narrow-band filter transmission, because the photometric redshifts are already well constrained for this sample and searching the full filter width would unnecessarily increase the false-positive rate without a corresponding gain in recovered detections. At each window position we compute the integrated flux significance within the aperture defined above, and the window yielding the maximum value is adopted as the spectral line centre of the Ly$\alpha$ emission.

For three sources (JELS IDs J100037.6+022012.6, J100040.4+022004.4, and J100027.3+022017.0), spectroscopic redshifts are available through the Dawn \textit{JWST} Archive \citep[DJA;][]{deGraaff2024, Heintz2024}, drawing on observations from \textit{JWST} programmes (GO 6585; \citealt{Coulter2024} and GO 6368; \citealt{Dickinson2024}). For these sources we fix the spectral search window to the spectroscopic redshift $\pm 35$\,\AA\ (the average search-window size in the sample is $\sim70$\,\AA) rather than scanning across $[z_\mathrm{min}, z_\mathrm{max}]$. The offsets between the JELS photometric and spectroscopic redshifts are $\Delta z = [0.005, 0.02, 0.004]$, corresponding to wavelength differences of [5.71\,\AA, 27.47\,\AA, 5.23\,\AA], respectively. All lie comfortably within the corresponding search windows. This agreement provides confidence that the photometric-redshift bounds are sufficiently broad to capture Ly$\alpha$ emission for the remainder of the sample and that we are unlikely to miss detections because of an overly restrictive redshift prior.

\subsubsection{Significance definition and false-positive calibration}

At each trial position and wavelength window we compute an integrated significance, $\Sigma_{\mathrm{Ly}\alpha}$, defined as the continuum-subtracted flux integrated within the adopted spatial aperture and spectral window divided by the corresponding propagated flux uncertainty. Although analogous to a conventional signal-to-noise ratio, $\Sigma_{\mathrm{Ly}\alpha}$ does not follow the Gaussian statistics expected for a fixed-aperture measurement because it is subsequently maximised over multiple spatial and spectral trials. We therefore characterise the distribution of $\Sigma_{\mathrm{Ly}\alpha}$ empirically through a false-positive analysis \citep[following the approach of][]{Guo2024}.

\begin{figure}
	\includegraphics[width=\columnwidth]{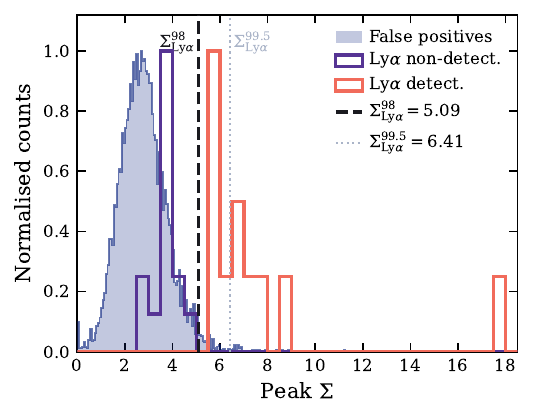}
	\caption{Peak $\Sigma_{\mathrm{Ly}\alpha}$ distributions for constructed false positive sources (filled) and JELS Ly$\alpha$ candidates (outlines). JELS candidates above the detection threshold (98th percentile: $\Sigma_{\mathrm{Ly}\alpha}^{98} = 5.09$) are shown in red and the candidates below this threshold are shown in purple. All distributions are normalised to their own peak. The $\Sigma_{\mathrm{Ly}\alpha}^{99.5}$ line is shown to demonstrate the threshold for strong detections.}
	\label{fig:snr_hist}
\end{figure}

For each continuum-subtracted subcube we evaluate 500 random spatial positions, excluding an $8\ \mathrm{arcsec}\times8\ \mathrm{arcsec}$ region centred on the JELS source and a $1\ \mathrm{arcsec}$ border around the edge of the subcube. Each position is treated as a potential false-positive source and passed through the identical spatial and spectral optimisation procedure described above, with the spectral search range set to the same $[z_\mathrm{min}, z_\mathrm{max}]$ interval as the corresponding real source. This ensures that the false-positive population samples the same parameter space and is subject to the same optimisation bias as the real detections. The resulting distribution of peak $\Sigma_{\mathrm{Ly}\alpha}$ values is shown in Figure~\ref{fig:snr_hist}. The median value of the false-positive population is $\Sigma_{\mathrm{Ly}\alpha} = 2.7$, reflecting the tendency of the optimisation procedure to select positive noise fluctuations. We therefore define our detection threshold at the 98th percentile of this empirical distribution, $\Sigma_{\mathrm{Ly}\alpha}^{98} = 5.09$. Sources exceeding this threshold are classified as detections, while those below it are treated as non-detections. We further define a \textit{strong} detection threshold at the 99.5th percentile, $\Sigma_{\mathrm{Ly}\alpha}^{99.5} = 6.41$. For a parent sample of 24 sources, these thresholds correspond to expected contamination rates of $\lesssim0.5$ and $\lesssim0.125$ false positives, respectively, providing a practical distinction between secure and highly secure detections.

\subsection{Ly{\fontsize{10}{10}\selectfont$\mathbf{\alpha}$} flux measurement}
\label{sec:lya_flux}

One-dimensional spectra are extracted from the continuum-subtracted MUSE datacubes using a circular aperture of radius $0.6\ \mathrm{arcsec}$, centred on the Ly$\alpha$ emission of each source. This aperture was chosen to maximise signal-to-noise of the Ly$\alpha$ emission as described in Section ~\ref{sec:cog}. The variance spectrum is extracted from the corresponding MUSE variance sub-cube by summing the per-spaxel variances within the aperture, and is used to weight the spectral fit and compute flux uncertainties.

Following \citet{Bacon2023}, we fit the Ly$\alpha$ line with a skewed Gaussian profile to account for the characteristic asymmetric red wing produced by resonant scattering in the ISM and attenuation by the IGM \citep{Verhamme2006, Mason2018}. We adopt the skew-normal parametrisation of \citet{Azzalini1999}:
\begin{equation}
    f(\lambda) = \frac{F_{\rm Ly\alpha}}{\sigma\sqrt{2\pi}}\,
    \exp\!\left(-\frac{(\lambda-\lambda_0)^2}{2\sigma^2}\right)
    \left[1 + \mathrm{erf}\!\left(\frac{\gamma(\lambda-\lambda_0)}{\sqrt{2}\,\sigma}\right)\right],
\end{equation}
where $\lambda_0$ is the central wavelength, $\sigma$ is the wavelength dispersion, $\gamma \geq 0$ is the asymmetry parameter controlling the extent of the red wing, and $\mathrm{erf}$ is the error function. When $\gamma = 0$ the profile reduces to a standard Gaussian. Because the profile integrates to unity, $F_{\rm Ly\alpha}$ directly gives the total integrated line flux, evaluated using the \textsc{SciPy} \textsc{skewnorm} implementation \citep{Virtanen2020}. The best-fit profile is obtained by minimising the variance-weighted residuals between the model and the extracted spectrum using \textsc{scipy.optimize.curve\_fit}, a bounded non-linear least-squares fit in which the per-spaxel variances provide the weights. The fit is performed within a $\pm25$\,\AA\ window centred on the Ly$\alpha$ line centre determined by the spectral window search in Section~\ref{sec:lyadetectsig}, with parameters bounded to physically motivated ranges. The asymmetry parameter is restricted to $\gamma \geq 0$, reflecting the expected red asymmetry of Ly$\alpha$ from resonant scattering in outflowing gas, with an upper bound of $\gamma = 8$ to prevent degenerate solutions. The scale parameter is bounded to $\sigma \geq 1.0$\,\AA, approximately corresponding to the instrumental dispersion of MUSE, preventing the fit from converging to unphysically narrow profiles driven by noise fluctuations.

For sources with a successful fit (detections), the integrated flux is taken directly from the best-fit $F_{\rm Ly\alpha}$. Rather than adopt the fit covariance, we estimate the uncertainty through a Monte Carlo resampling of the fit. Each source is refit 500 times, with each realisation generated by perturbing the best-fit model by noise drawn from the propagated variance spectrum and refitting under the same model, bounds and variance weighting as the science run. The flux uncertainty is taken as the standard deviation of the recovered $F_{\rm Ly\alpha}$ distribution, which avoids the tendency of covariance-matrix uncertainties to underestimate the true error. The line width is recorded as the FWHM computed numerically from the model profile in km\,s$^{-1}$. As a consistency check, we compared these profile-fit fluxes against a direct integration of the continuum-subtracted spectrum within the line window. The two methods agree on a 1:1 basis within the flux uncertainties, confirming that the skewed-Gaussian fit recovers the full line flux.

For sources that do not exceed the adopted detection threshold, we compute conservative $5\sigma$ flux upper limits using a fixed integration window centred on the Ly$\alpha$ wavelength corresponding to the JELS photometric redshift. The choice of a $5\sigma$ upper limit is broadly comparable to the empirical detection threshold ($\Sigma \approx 5$) adopted in Section~\ref{sec:lyadetectsig}, ensuring that reported upper limits and detections are defined at similar significance levels. The window width is scaled to the median FWHM of the detected population, 278 km\,s$^{-1}$, ensuring upper limits are computed on a consistent basis across the full 
sample:
\begin{equation}
    F_{\rm lim} = 5  \sigma_{\rm rms} \sqrt{N_{\rm pix}}  \Delta\lambda,
\end{equation}
where $\sigma_{\rm rms}$ is the per-pixel noise estimated from the propagated variance spectrum within the integration window, $N_{\rm pix} = \mathrm{FWHM}/\Delta\lambda$ is the effective number of spectral pixels, and $\Delta\lambda$ is the spectral pixel scale. Representative examples of the spectral fits and pseudo-narrow-band images are shown in Figure \ref{fig:detect}.

\subsection{Ly{\fontsize{10}{10}\selectfont$\mathbf{\alpha}$} Equivalent Widths}
\label{sec:ew}
Rest-frame Ly$\alpha$ equivalent widths (EWs) are computed as the ratio of the integrated Ly$\alpha$ line flux to the UV continuum flux density evaluated at the observed Ly$\alpha$ wavelength, divided by $(1+z)$ to convert to the rest frame. The MUSE continuum is not detected at sufficient significance to measure this directly for most sources, so the UV continuum flux density at the observed Ly$\alpha$ wavelength is instead estimated by extrapolating from the rest-frame UV using the power-law continuum slope $\beta$ (where $f_\lambda \propto \lambda^\beta$) and the absolute UV magnitude $M_{\rm UV}$, both measured directly from a power-law fit to the rest-frame UV photometry (Section~\ref{sec:sed}). For sources where the UV slope is moderately well constrained ($\sigma_\beta < 0.5$), we use the power-law fitted $\beta$ directly. For sources with poorly constrained slopes ($\sigma_\beta \geq 0.5$), we instead anchor the continuum to the PRIMER F150W broadband flux density, which samples the rest-frame UV at this redshift, and extrapolate to the observed Ly$\alpha$ wavelength using a fixed $\beta = -1.92 \pm 0.05$. This value is the direct measurement for the parent JELS H$\alpha$-selected sample from \citet{Pirie2025}. For these fixed-slope sources we propagate the $\pm 0.05$ uncertainty on $\beta$ into the continuum estimate. Of Ly$\alpha$ detections for which EWs are measured, half of the sources require this fixed-slope treatment. This relatively high fraction reflects the intrinsic UV faintness of the JELS H$\alpha$-selected sample. This faintness is an advantage in other respects, since the sample probes the low-luminosity population that dominates the ionizing budget during reionization, but it does limit the precision to which the UV continuum, and hence the EW, can be constrained for part of the sample. However, the only result we draw from the EWs is the correlation with $f_\mathrm{esc}^{\mathrm{Ly}\alpha}$ discussed in Section~\ref{sec:proprelations}, which we assess with a rank correlation statistic. Varying the fixed $\beta$ value shifts individual EWs only within their uncertainties and leaves the rank order of the sample unchanged, so this correlation is insensitive to the adopted slope. EW uncertainties are propagated in quadrature from the uncertainties on both the line flux and the continuum flux density estimate.

\section{Ly{\fontsize{14}{14}\selectfont$\mathbf{\alpha}$} properties of $\mathbf{z\approx6}$ H{\fontsize{14}{14}\selectfont$\mathbf{\alpha}$}-emitters}
\label{sec:results}
\begin{figure*}
    \centering
	\includegraphics[width=\textwidth]{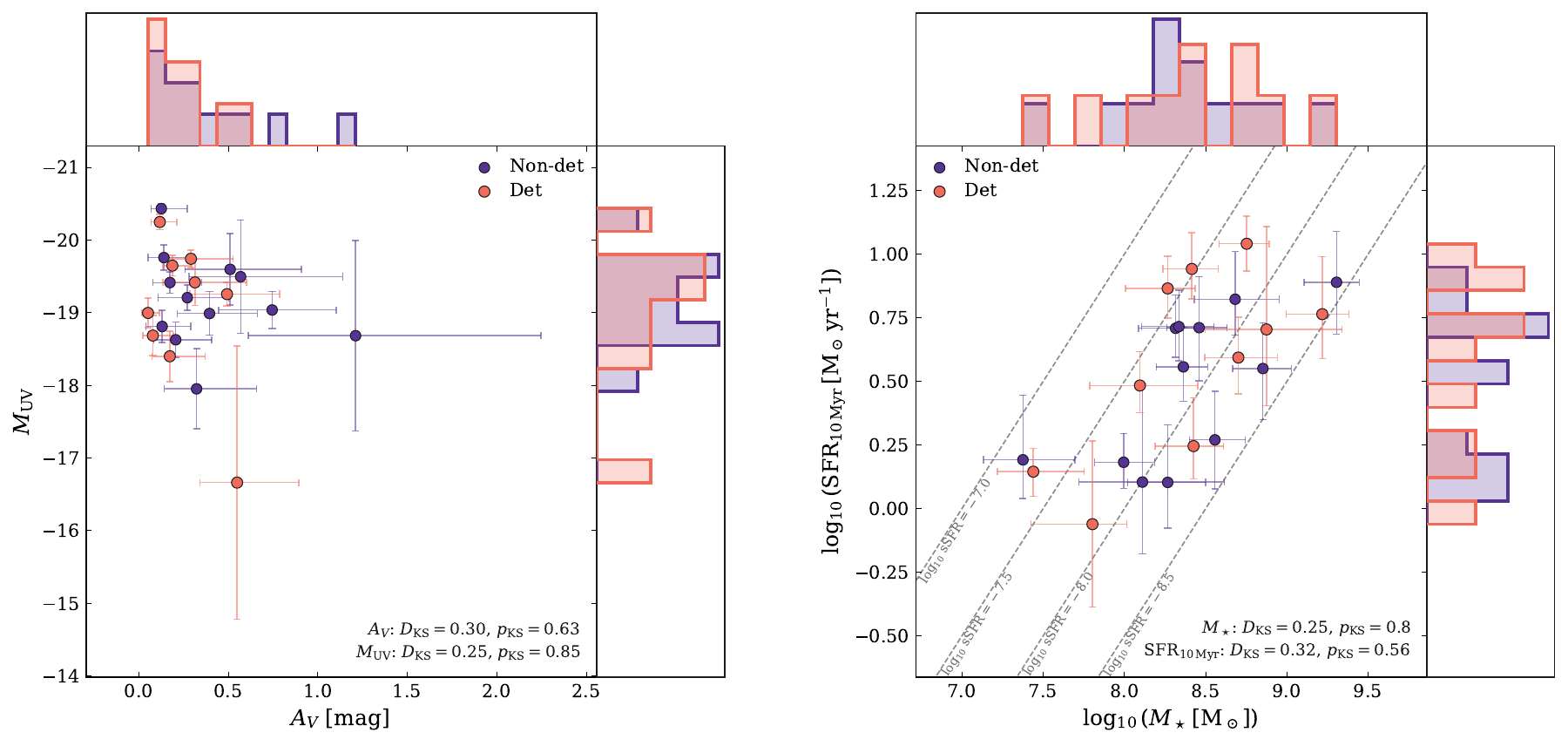}
    \caption{Distributions of dust attenuation, UV magnitude, stellar mass, and star-formation rate for Ly$\alpha$-detected (red) and Ly$\alpha$-undetected (purple) H$\alpha$ emitters at $z \sim 6$. The left panel shows $A_V$ against $M_{\mathrm{UV}}$ and the right panel shows SFR$_{10\,\mathrm{Myr}}$ against $M_\star$, with lines of constant sSFR drawn on the right panel. The distribution of each property is shown as a histogram alongside the corresponding axis, normalised to the total number of sources in its respective subsample. Annotated $D_{KS}$ and $p_{KS}$ values are the median of a bootstrapped two-sample Kolmogorov-Smirnov test, in which each source is perturbed by its measurement uncertainty over 2000 realisations. A low $p_{KS}$ would indicate that the two distributions are inconsistent with being drawn from the same parent population. AGN candidates are excluded from all panels.}
	\label{fig:det_hist}
\end{figure*}

Using our detection threshold $\Sigma_{\mathrm{Ly}\alpha}$ from Section~\ref{sec:lyadetectsig}, we find 12 of the 24 sources to be above $\Sigma_{\mathrm{Ly}\alpha}^{98}$, giving an overall Ly$\alpha$ detection fraction of $50 \pm 10$ per cent (12/24 sources), where the uncertainty is binomial. Of these 12 detections, 7 exceed $\Sigma_{\mathrm{Ly}\alpha}^{99.5}$, making them our high-confidence detections. The spectra and narrow-band cutouts of all detections are shown in Figure~\ref{fig:detect}. The 12 sources below $\Sigma_{\mathrm{Ly}\alpha}^{98}$ are classified as non-detections and are presented in Figure~\ref{fig:nondetect}. The following analysis adopts upper limits on the Ly$\alpha$ flux for all non-detections, as calculated in Section~\ref{sec:lya_flux}.

Many studies of Ly$\alpha$ emission at high redshift define Ly$\alpha$ emitters (LAEs) using a threshold of $\rm{EW}(\mathrm{Ly}\alpha) > 25\,\text{\AA}$ \citep[e.g.][]{Stark2010, Stark2011, Schenker2014, Kusakabe2020}. To compare like-for-like with the UV-faint Lyman-break galaxy (LBG) samples of \citet{Stark2011}, we restrict this measurement to their fiducial magnitude range $-20.25 < M_{\rm UV} < -18.75$, within which 16 of our sources fall. We further account for sensitivity, since for a small number of non-detections the local detection threshold $\Sigma_{\mathrm{Ly}\alpha}^{98}$ lies above the Ly$\alpha$ flux expected for an $\rm{EW}({\rm Ly\alpha}) = 25$\,\AA\ line at the source redshift and continuum level. A galaxy at exactly the canonical LAE threshold would not have been recovered in these cases, so they cannot be counted as fair non-detections of LAEs. One such source falls within the adopted $M_{\rm UV}$ range, leaving 15 sources for which a canonical LAE would have been detectable. Of these, 5 are LAEs, giving $X_{\rm Ly\alpha} = 33 \pm 12$ per cent (5/15 sources). This is lower than the $54 \pm 11$ per cent reported by \citet{Stark2010, Stark2011} for UV-faint LBGs at $z\sim6$, though the two remain consistent within the binomial uncertainties. More recent studies have generally found lower Ly$\alpha$ fractions than these earlier DEIMOS/Keck measurements. \citet{Jones2025} show that recent \textit{JWST}-era measurements at $4<z<6.5$ are systematically lower than the values reported by \citet{Stark2010, Stark2011}$,$ while remaining consistent with other recent spectroscopic studies, including \citet{Napolitano2024} and \citet{Tang2024a}. Part of this apparent decline may be instrumental rather than astrophysical. The narrow NIRSpec micro-shutter assembly (MSA) slitlets ($\sim 0.2\arcsec \times 0.46\arcsec$) can incur substantial slit losses for Ly$\alpha$ emission that is spatially extended or offset from the UV continuum, whereas the wider Keck slits and our own MUSE apertures are more forgiving. \citet{Jiang2024} provide an extreme example, where a source with MUSE-detected Ly$\alpha$ at $\rm{EW} \sim 75$\,\AA\ has no detected Ly$\alpha$ in \textit{JWST}/NIRSpec. Our measured LAE fraction therefore lies comfortably within the range reported by recent surveys at similar redshifts.

Recent studies have also highlighted substantial scatter in the measured Ly$\alpha$ fraction at fixed redshift, driven by differences in sample selection, completeness corrections, survey depth, and cosmic variance \citep[e.g.][]{Kusakabe2020, Napolitano2024, Jones2025}. In particular, \citet{Napolitano2024} demonstrate that field-to-field variations alone can produce a wide range of observed Ly$\alpha$ fractions at high redshift, consistent with expectations from inhomogeneous reionization models. Using CoDa II reionization simulations, they find a $2\sigma$ interval of $X_{\mathrm{Ly}\alpha}=[0.12,0.48]$ for $\rm{EW}(\mathrm{Ly}\alpha)>25\text{\,\AA}$, and our $X_{\mathrm{Ly}\alpha}$ is fully encompassed in this range. Similarly, \citet{Jones2025} show that completeness corrections and selection effects can systematically alter inferred Ly$\alpha$ fractions, particularly for low-EW emitters and UV-faint galaxies. As our sample is drawn from a single field, we cannot rule out that cosmic variance in the IGM has shifted our measured $X_{\mathrm{Ly}\alpha}$ within this range, though this same restriction to one field is what allows us to isolate intrinsic scatter from field-to-field IGM effects, as discussed further in Section~\ref{sec:igm_uniformity}.

Our slightly lower value relative to the original LBG studies may nevertheless indicate the onset of mild IGM suppression. Our sample lies at $z\sim6.1$, where several works have reported evidence for a gradual decline in the Ly$\alpha$ fraction beyond $z\gtrsim6$ \citep[e.g.][]{Pentericci2011, Schenker2014, Mason2018, Jones2025}. \citet{Kusakabe2020}, using deep MUSE observations extending to significantly fainter UV magnitudes than earlier LBG studies, also found systematically lower Ly$\alpha$ fractions consistent with a possible turnover beginning around $z\sim5.5$. They further argued that UV-selected LBG samples may preferentially include strong Ly$\alpha$ emitters due to photometric selection biases. Our H$\alpha$-selected sample avoids this bias, providing a more representative census of the star-forming galaxy population at this epoch, down to our H$\alpha$ flux limit.

The detected sources span Ly$\alpha$ luminosities of $9.0 \times 10^{41}$ to $4.8 \times 10^{42}\ \mathrm{erg\ s^{-1}}$, with a median of $1.6 \times 10^{42}\ \mathrm{erg\ s^{-1}}$. The median rest-frame Ly$\alpha$ equivalent width of the detections is 48\,\AA, consistent with the faint star-forming galaxy population at $z\sim6$ \citep[e.g.][]{Kusakabe2020, Saxena2024}. The skewed Gaussian fits (Section~\ref{sec:lya_flux}) yield a mean asymmetry parameter of $\bar{\gamma}=4.7$ and a median of $\tilde{\gamma}=5.5$, with values ranging from near-symmetric to highly asymmetric profiles, consistent with the diversity of Ly$\alpha$ line morphologies observed in faint high-redshift galaxies \citep[e.g.][]{Bacon2023}. For the brightest detections this asymmetry is clearly visible by eye in the spectra shown in Figure~\ref{fig:detect}. The Ly$\alpha$ velocity offset, which requires systemic redshifts from rest-frame optical emission lines, will be explored in future work using the full JELS-MUSE dataset. In the following sections we investigate how the inferred $f_{\rm esc}^{\rm Ly\alpha}$ varies across the observed range of galaxy properties in our sample.

\subsection{Drivers of Ly{\fontsize{10}{10}\selectfont$\mathbf{\alpha}$} detectability}
\label{sec:drive}

We now examine the properties distinguishing detections from non-detections to investigate whether the absence of Ly$\alpha$ emission is driven by observational limitations or by intrinsic galaxy properties regulating Ly$\alpha$ escape. Figure~\ref{fig:det_hist} shows the two populations in the $A_V$ against $M_{\mathrm{UV}}$ plane and the SFR$_{10\,\mathrm{Myr}}$ against $M_\star$ plane, with histograms of the distribution of each property shown alongside. For every property we quantify the separation between detections and non-detections with a two-sample KS test. To account for the measurement uncertainties, which are largest for the faintest sources, we perform each test in a bootstrapped manner. In each of 2000 realisations we perturb the value of every source by its measurement uncertainty,
drawing from a split-normal distribution to respect the asymmetric posteriors where present, and recompute the KS statistic on the perturbed samples. We report the median $D_{KS}$ and $p_{KS}$ across these realisations.

We first test whether non-detections are preferentially faint. Comparing rest-frame UV absolute magnitudes, $M_{\mathrm{UV}}$, for detections and non-detections, we find no statistically significant difference ($D_{KS} = 0.25$, $p_{KS} = 0.85$), with only 2 per cent of realisations falling below $p_{KS} = 0.05$. The non-detections therefore reflect genuine variations in Ly$\alpha$ escape rather than insufficient sensitivity, and our sample is not significantly biased by the MUSE sensitivity or by the faintness of the population. There are, however, a small number of sources (3 of 24) where the depth does not reach the flux of a canonical $\rm{EW} = 25$\,\AA\ emitter. This is relevant when comparing to studies adopting this threshold, though our sensitivity extends to lower equivalent widths across the rest of the sample (see introduction to Section~\ref{sec:results}).

All sources occupy the same field within a narrow redshift range, so large object-to-object variations in IGM transmission are not expected. We test this explicitly in Section~\ref{sec:igm_uniformity} using reionization simulations, and confirm that the expected spread in transmission across a field of this size is smaller than the scatter we measure in $f_{\mathrm{esc}}^{\mathrm{Ly}\alpha}$. At fixed redshift, scatter in Ly$\alpha$ visibility is dominated by ISM conditions \citep{Dijkstra2014, Dijkstra2017}. We therefore focus on intrinsic galaxy properties. Dust represents a natural candidate as resonant scattering increases Ly$\alpha$ path length and thus susceptibility to dust absorption. We use V-band attenuation ($A_V$) derived from SED fitting as our dust tracer, which is better constrained than the UV continuum slope $\beta$ for a significant subset of our sources where the rest-UV photometry is limited. We find no statistically significant difference in $A_V$ between detections and non-detections ($D_{KS} = 0.30$, $p_{KS} = 0.63$).

We further examine star-formation rates averaged over the past 10\,Myr (SFR$_{10\,\mathrm{Myr}}$), taken from the BAGPIPES SED fits (Section~\ref{sec:sed}), which trace the recent ionizing photon production and feedback activity most directly relevant to Ly$\alpha$ escape, along with the stellar mass $M_\star$. Again, no clear separation between the two populations is observed ($D_{KS} = 0.25$, $p_{KS} = 0.80$ for $M_\star$ and $D_{KS} = 0.32$, $p_{KS} = 0.56$ for SFR$_{10\,\mathrm{Myr}}$). We note that the sample sizes in this analysis are small, and the KS statistic values range between $D_{KS} = 0.25$ and $0.32$ across these tests, so the number of sources would need to be increased to place tighter constraints. Nevertheless, the consistently high p-values give us confidence that there is no significant systematic difference between the detection and non-detection populations in any of the properties examined. The global galaxy properties considered here are therefore not the primary factors governing Ly$\alpha$ escape in our sample.

The absence of trends in any of these diagnostics points to the complexity of Ly$\alpha$ radiative transfer. Escape is governed by the detailed distribution, kinematics, and covering fraction of neutral hydrogen on scales not probed by the observables considered here \citep{Gazagnes2020, Mauerhofer2021, Saldana-Lopez2022, Jaskot2025, Flury2025}. Models in which low-column-density channels punctuate an otherwise opaque medium \citep[e.g. the `picket-fence';][]{Heckman2001, Reddy2016} produce a highly anisotropic, stochastic Ly$\alpha$ escape that depends strongly on viewing angle \citep{Behrens2014, Cen2015, Gronke2016, Kimm2019, Jaskot2019}. Stellar feedback and outflows can additionally Doppler-shift Ly$\alpha$ photons out of resonance, providing a further, kinematic route to escape independent of covering fraction \citep{Verhamme2006, Dijkstra2014, Flury2025}. Consistent with this picture, \citet{Flury2022b} argued that geometry and orientation effects may contribute substantially to the observed scatter in both Ly$\alpha$ and LyC escape. Together, these effects mean that small variations in neutral gas geometry, orientation, and kinematics can drive large changes in Ly$\alpha$ visibility, rendering broad photometric diagnostics poor predictors of detectability in individual systems. This complexity underscores the value of large statistical samples, which can help average over object-to-object variations and reveal underlying population trends. Nevertheless, substantial scatter can persist even in large surveys \citep[e.g.][]{Flury2022b}, indicating that geometry, orientation, and neutral gas structure remain important drivers of Ly$\alpha$ and LyC escape across a wide range of galaxy populations.

\section{Ly{\fontsize{14}{14}\selectfont$\mathbf{\alpha}$} escape in the JELS H{\fontsize{14}{14}\selectfont$\mathbf{\alpha}$} population}
\label{sec:lyaesc}

\begin{figure*}
    \centering
	\includegraphics[width=0.85\textwidth]{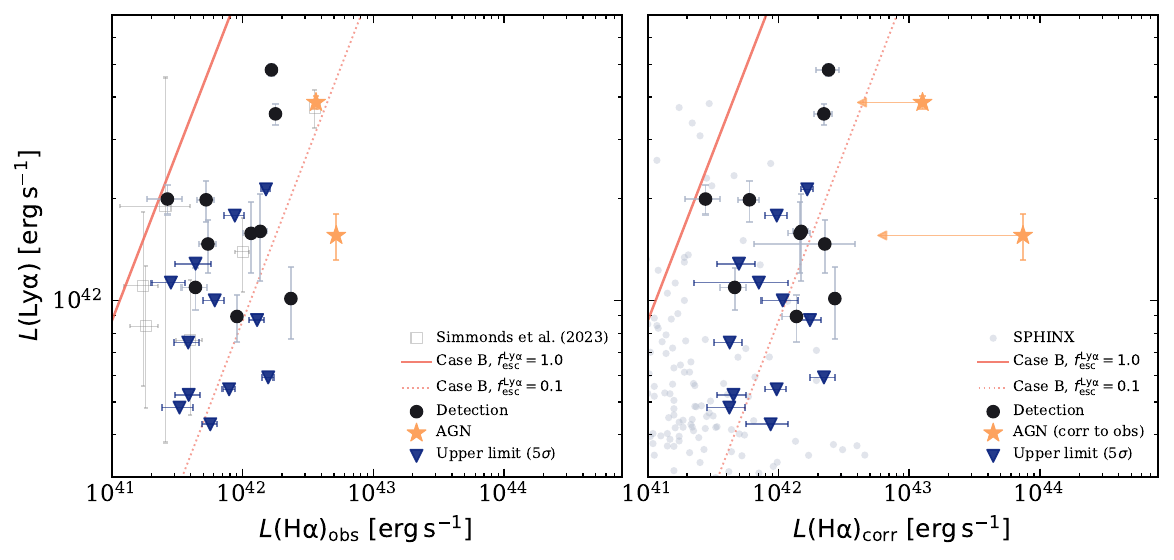}
	\caption{
    Comparison of Ly$\alpha$ and H$\alpha$ luminosities.
    Left: observed Ly$\alpha$ luminosity as a function of observed H$\alpha$ luminosity with the H$\alpha$ luminosities aperture corrected but not dust corrected. The grey points show the $z > 6$ sample from \citet{Simmonds2023}.
    Right: Ly$\alpha$ luminosity as a function of dust-corrected H$\alpha$ luminosity. The background density distribution shows galaxies from the SPHINX simulations \citep{Rosdahl2018, Rosdahl2022, Katz2023}. In both panels, the red solid line indicates the expected relation between Ly$\alpha$ and H$\alpha$ luminosities assuming Case B recombination and $f_{\rm esc}^{\rm Ly\alpha} = 1$, and the dashed line indicates $f_{\rm esc}^{\rm Ly\alpha} = 0.1$. Filled black circles represent detections and filled blue triangles indicate upper limits. AGN candidates are marked by yellow stars. In the right panel, arrows connect these sources to their dust-uncorrected positions, as the SED fitting does not include AGN components and may therefore yield unreliable dust correction estimates for these systems.
    }
	\label{fig:lya_ha}
\end{figure*}

The Ly$\alpha$ escape fraction, $f_{\rm esc}^{\rm Ly\alpha}$, quantifies the fraction of Ly$\alpha$ photons produced in \ion{H}{ii} regions that escape the galaxy without being absorbed by dust or scattered out of the line-of-sight by neutral hydrogen. Under Case B recombination, atomic physics predicts an intrinsic Ly$\alpha$-to-H$\alpha$ luminosity ratio of $\simeq 8.7$ for typical nebular conditions \citep{Storey1995}, with a weak dependence on temperature and density at the $\sim$5 per cent level. We therefore adopt this standard value throughout and compute

\begin{equation}
    f_{\rm esc}^{\rm Ly\alpha} = \frac{1}{8.7} \cdot 
    \frac{L_{\rm Ly\alpha}}{L_{\rm H\alpha}},
    \label{eq:fesc}
\end{equation}

\noindent where $L_{\rm Ly\alpha}$ is the observed Ly$\alpha$ luminosity from the MUSE spectra (Section~\ref{sec:lya_flux}) and $L_{\rm H\alpha}$ is the dust-corrected H$\alpha$ luminosity from the JELS narrow-band photometry (see Section~\ref{sec:haflux}). For sources undetected in Ly$\alpha$, we derive upper limits on $f_{\rm esc}^{\rm Ly\alpha}$ using the $5\sigma$ Ly$\alpha$ flux limits described in Section~\ref{sec:lya_flux}. Uncertainties are propagated in quadrature from the uncertainties on both luminosities.

We highlight that $L_{\rm Ly\alpha}$ is intentionally left as the \textit{observed}, dust-uncorrected quantity, following standard practice for the calculation of $f_{\rm esc}^{\rm Ly\alpha}$ \citep[e.g.][]{ Simmonds2023, Lin2024, Shimizu2026}. Unlike H$\alpha$, where dust extinction can be estimated from the UV-optical SED, Ly$\alpha$ photons are subject to both dust absorption and resonant scattering by neutral hydrogen in the ISM and CGM, and these two loss channels cannot be separated observationally. Dust correction of the observed Ly$\alpha$ flux would therefore not recover the intrinsic recombination luminosity as it would only partially account for one of the two mechanisms that suppress escape. More fundamentally, $f_{\rm esc}^{\rm Ly\alpha}$ as defined here measures the fraction of ionizing-photon-traced Ly$\alpha$ production that survives to escape the galaxy entirely, whether lost to dust or scattered out of the line-of-sight. This is the physically meaningful quantity for reionization studies, where the IGM is sensitive to the photons that escape the galaxy's ISM and CGM and not to those absorbed internally.

Figure~\ref{fig:lya_ha} shows the observed Ly$\alpha$ luminosities against H$\alpha$ luminosities for our sample, split into two panels to reflect different assumptions about dust attenuation. In the left panel, we plot against the uncorrected H$\alpha$ luminosities alongside the \citet{Simmonds2023} comparison sample at similar redshift ($z>6$). As \citet{Simmonds2023} note, their sample is consistent with little to no dust attenuation, making the uncorrected H$\alpha$ a reasonable tracer of intrinsic recombination emission for those galaxies. For our sample, however, SED fitting indicates that although dust attenuation is low, it is non-negligible and can be significant for some sources. 
The presence of dust means that the uncorrected H$\alpha$ luminosity therefore underestimates the true ionizing photon production rate, and consequently overestimates $f_{\rm esc}^{\rm Ly\alpha}$. In the right panel we therefore plot against dust-corrected H$\alpha$ luminosities, which better represents the intrinsic recombination luminosity required by Equation~\ref{eq:fesc}. The solid red line in both panels shows the Case B expectation of $L_{\rm Ly\alpha} = 8.7\,L_{\rm H\alpha}$ \citep{Storey1995} and the dashed line shows $f_{\rm esc}^{\rm Ly\alpha}$ = 10 per cent.

To compare our observed galaxies with theoretical expectations, we include in the right panel galaxies from the public $\mathrm{SPHINX}^{20}$ cosmological radiation-hydrodynamics simulation data release \citep{Katz2023}, based on the simulations of \citet{Rosdahl2018, Rosdahl2022}, that model high-redshift galaxy formation including stellar feedback, ionizing radiation transfer, dust attenuation, and Ly$\alpha$ radiative transfer. We use the $z = 6$ \textsc{SPHINX} catalogue, taking intrinsic H$\alpha$ luminosities from the public data release, which represent the recombination luminosity prior to dust attenuation as computed directly from the simulated nebular gas via photoionization modelling. Observed Ly$\alpha$ luminosities are obtained from the \textsc{SPHINX} Ly$\alpha$ radiative transfer products, where photons are propagated through neutral hydrogen and dust via Monte Carlo radiative transfer accounting for resonant scattering, absorption, and anisotropic escape. We use the escaped luminosity averaged over the ten provided viewing angles. This pairing of intrinsic H$\alpha$ and escaped Ly$\alpha$ luminosities makes the \textsc{SPHINX} points directly comparable to the dust-corrected observational quantities in the right panel and to our $f_{\rm esc}^{\rm Ly\alpha}$ definition in Equation~\ref{eq:fesc}. We note that the \textsc{SPHINX} volume contains relatively few galaxies at the bright, high-mass end, so there is limited overlap in parameter space between the simulation and our sample at the luminosities probed here (Figure~\ref{fig:lya_ha}). The \textsc{SPHINX} points are therefore included as a broad theoretical reference rather than for a source-by-source comparison.

We also note that \citet{Choustikov2024} showed that the standard observational method of inferring $f_{\rm esc}^{\rm Ly\alpha}$ from dust-corrected H$\alpha$ tends to over-predict the true Ly$\alpha$ escape fraction by as much as two orders of magnitude, with a mean absolute error of $0.43$~dex, driven primarily by the failure of dust corrections applied to sight-line-attenuated H$\alpha$ to fully recover the intrinsic emission. By using the intrinsic H$\alpha$ directly from the simulation, the \textsc{SPHINX} escape fractions shown here are not themselves subject to this error. Our observed escape fractions, however, are still derived from dust-corrected H$\alpha$ and so are subject in principle to a systematic of this kind. We do not apply the \citet{Choustikov2024} value as a correction, since their analysis assumes a somewhat different measurement method to ours, and we quote it only to indicate the likely direction and approximate scale of the effect. It should therefore be considered when comparing the observed and simulated escape fractions rather than treated as a calibrated offset. A residual systematic remains from the fixed intrinsic Ly$\alpha$-to-H$\alpha$ ratio of 8.7, which has a weak dependence on electron temperature and density not captured by a single canonical value, but this is expected to be small compared to the dominant dust correction uncertainty \citep{Choustikov2024}.

\subsection{How uniform do we expect IGM conditions to be?}
\label{sec:igm_uniformity}

Interpreting the scatter in $f_{\mathrm{esc}}^{\mathrm{Ly}\alpha}$ requires some understanding of how uniform the IGM conditions are likely to be across the sample. The observed escape fraction can be thought of as the product of an intrinsic, ISM and CGM driven term and the IGM transmission of Ly$\alpha$, so galaxy-to-galaxy variation in line-of-sight transmission would in principle contribute to the spread we measure. Our sources lie in a single contiguous field across a narrow redshift interval, so we might expect these differences to be relatively small, but it is worth testing this expectation directly.

\begin{figure*}
	\includegraphics[width=0.9\textwidth]{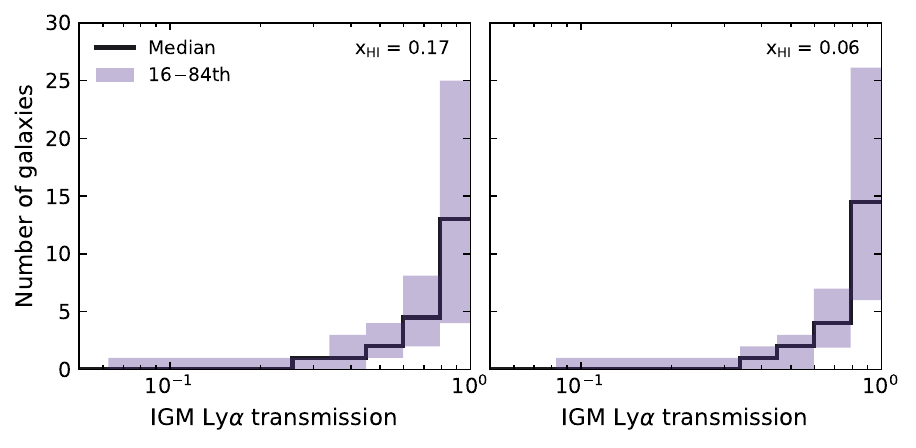}
	\caption{Distribution of IGM Ly$\alpha$ transmission, $\mathcal{T}_{\mathrm{IGM}}$ (Equation~\ref{eq:tigm}), for galaxies in the mock survey regions extracted from the reionization simulation at $z = 6$. The two panels show the two neutral-fraction models that bracket the global value at this redshift, $x_{\mathrm{HI}} = 0.17$ (left) and $x_{\mathrm{HI}} = 0.06$ (right). In each panel the solid line shows the median number of galaxies per transmission bin across the 144 mock regions, and the shaded band shows the 16th to 84th percentile range across regions. In both models the transmission is strongly skewed towards unity, indicating that most galaxies sit in ionized regions and transmit the majority of their Ly$\alpha$. The scatter in transmission between galaxies is smaller than the spread we measure in $f_{\mathrm{esc}}^{\mathrm{Ly}\alpha}$, supporting the argument that IGM variation across a field of this size does not account for the observed range of escape fractions.}
	\label{fig:igm_transmission}
\end{figure*}

To do so we calculate the IGM transmission from two simulations of reionization (Keating et al., in prep). These simulations are performed with the radiative transfer code \textsc{aton} \citep{Aubert2008} in post-processing on top of a $160\,h^{-1}\,\mathrm{Mpc}$ cosmological hydrodynamic simulation performed with \textsc{p-gadget-3} \citep[last described in][]{Springel2005} as part of the Sherwood-Relics simulation suite \citep{Puchwein2022}. The simulation setup is similar to the description in \citet{Keating2020}. We consider two models that bracket the global neutral fraction at this redshift, $x_{\mathrm{HI}} = 0.17$ and $x_{\mathrm{HI}} = 0.06$. Both are consistent with an incomplete reionization at $z \approx 6$ as preferred by the Ly$\alpha$ forest \citep{Bosman2022}, and the true neutral fraction is expected to lie between them.
From the simulation outputs, we extract 144 mock survey regions at $z = 6$, each matched to the area and depth of our sample and spanning a redshift interval of $\Delta z \approx 0.1$. Galaxies are selected by a cut on halo mass ($M_{\mathrm{halo}} > 1.4 \times 10^{10}\,\mathrm{M}_\odot$), chosen to recover a median of 24 galaxies over all regions and so reproduce the size of our sample on average. For each galaxy we compute the IGM transmission of Ly$\alpha$ following the definition of \citet{Mesinger2015},

\begin{equation}
\mathcal{T}_{\mathrm{IGM}} = \frac{\int J(v)\, e^{-\tau_{\mathrm{Ly}\alpha}(v)}\, \mathrm{d}v}{\int J(v)\, \mathrm{d}v},
\label{eq:tigm}
\end{equation}

\noindent where $J(v)$ is the intrinsic Ly$\alpha$ emission profile and $\tau_{\mathrm{Ly}\alpha}(v)$ is the IGM optical depth along the line-of-sight. Results for the two reionization models are shown in the two panels of Figure~\ref{fig:igm_transmission}.

A few caveats apply to this setup. The galaxies are selected by a halo mass cut rather than by a reproduction of our H$\alpha$ selection, the boundary between the IGM and CGM is not sharply defined and is taken here as the virial radius, and the simulated volume ($160\,h^{-1}\,\mathrm{Mpc}$) is modest by reionization standards, so the bubble-size distribution may differ in larger volumes \citep{Zier2026}. These introduce some uncertainty into the predicted transmission distribution, including its width. Our argument does not rely on the precise value of that width, however, only on the transmission scatter being smaller than the spread we measure in $f_{\mathrm{esc}}^{\mathrm{Ly}\alpha}$.

In both models the transmission is strongly skewed towards high values, as expected near the end of reionization where most galaxies sit in ionized regions and transmit the majority of their Ly$\alpha$. This is in broad agreement with studies of the reionization bubble-size distribution, which find that galaxies can reside in ionized bubbles, including small ones, even when the global neutral fraction is low \citep{Lu2024, Neyer2024}. Some scatter in transmission remains between galaxies within a single region, so the IGM conditions are not identical across the sample. This scatter is, however, smaller than the spread we measure in $f_{\mathrm{esc}}^{\mathrm{Ly}\alpha}$. The galaxies in a field of this size are therefore expected to experience broadly similar IGM transmission, and differences in the IGM are unlikely to account for the full range of escape fractions we observe. Much of that range is instead likely to be intrinsic to the galaxies. A sample such as this, spanning little IGM variation, still shows substantial scatter in $f_{\mathrm{esc}}^{\mathrm{Ly}\alpha}$, which suggests that scatter measured at higher redshift should not be attributed to IGM patchiness alone, since a comparable intrinsic component is likely to be present.

However, we also note that not all of the remaining scatter can be attributed to differences in ISM and CGM properties. Part of it may reflect the relative velocity between each galaxy and its surrounding gas. Infalling gas is blue-shifted with respect to the galaxy and preferentially absorbs the blue side of the Ly$\alpha$ line, so galaxies with stronger infall transmit less of their Ly$\alpha$ even when the surrounding IGM is comparably ionized \citep{Sadoun2017, Mason2018, Weinberger2018}. The size of this effect depends on the intrinsic Ly$\alpha$ velocity offset, which is difficult to constrain without systemic redshifts and which we cannot measure for our sample. In the simulations this offset is fixed to reproduce the lowest-redshift results of \citet{Tang2024}, though the true value may differ. This does not change the picture above, since the IGM transmission across the sample remains broadly uniform in either case. It does mean that the intrinsic scatter need not stem from ISM and CGM property differences alone, and that the local velocity field around each galaxy may also play a part.

Taken together, these results suggest that our sample provides a useful reference point for intrinsic Ly$\alpha$ escape near the end of reionization. The scatter characterised here should be considered when interpreting measurements at higher redshift that are more heavily affected by the IGM, where suppression by a more neutral medium is superimposed on this same underlying variance.

\subsection{Average Ly{\fontsize{10}{10}\selectfont$\mathbf{\alpha}$} Escape Fraction and Redshift Evolution}
\label{sec:fesc_evolution}

\begin{figure*}
	\includegraphics[width=0.75\textwidth]{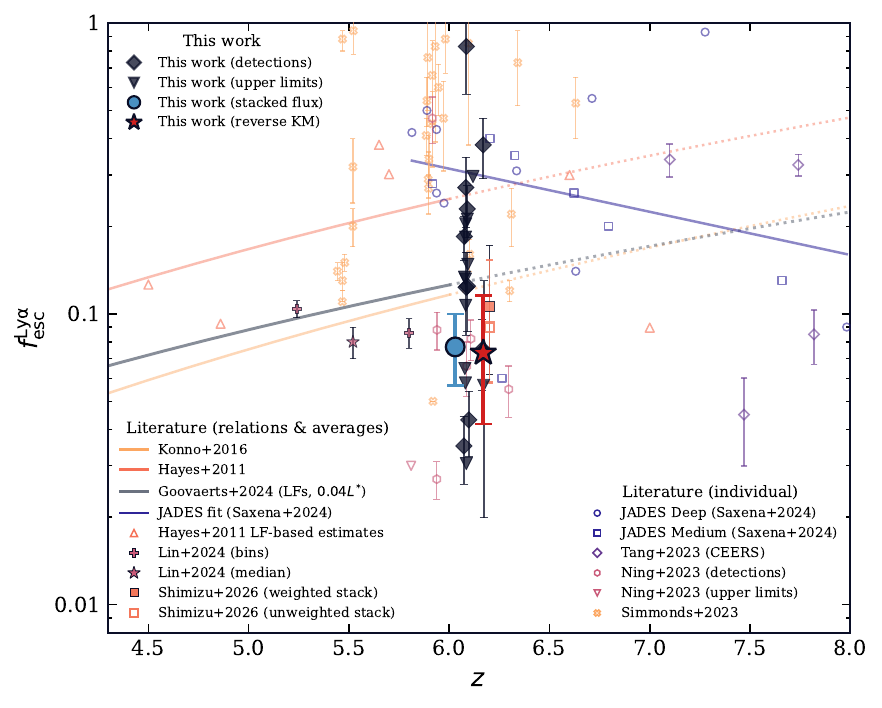}
	\caption{Ly$\alpha$ escape fraction as a function of redshift. Individual galaxies from the JELS-MUSE sample are shown as dark diamonds for detections and dark downward triangles for $5\sigma$ upper limits. The two population-averaged measurements derived in this work are shown as a blue circle (stacked flux) and a red star (reverse Kaplan-Meier median), both at $z \sim 6.1$ but slightly offset on the redshift axis for clarity. Literature relations and averages are shown in the lower-left legend: the relation of \citet{Konno2016}, the relation of \citet{Hayes2011} together with the underlying luminosity-function-based estimates from which it is derived, the luminosity-function estimate of \citet{Goovaerts2024} evaluated at $0.04\,L^\ast$, the JADES fit of \citet{Saxena2024}, the binned and median measurements of \citet{Lin2024}, and the weighted and unweighted stacks of \citet{Shimizu2026}. Solid lines indicate the redshift range constrained by each study and dotted extensions show extrapolation beyond it. Individual literature measurements are shown as faint open markers in the lower-right legend: the JADES Deep and Medium samples of \citet{Saxena2024}, the CEERS sample of \citet{Tang2023}, the detections and upper limits of \citet{Ning2023}, and the sample of \citet{Simmonds2023}.}
	\label{fig:fesc_redshift}
\end{figure*}

The Ly$\alpha$ escape fractions derived from the observed Ly$\alpha$-to-H$\alpha$ luminosity ratio (Equation~\ref{eq:fesc}) span a wide range across the sample. Individual detections range from $f_{\mathrm{esc}}^{\mathrm{Ly}\alpha} = 0.04$ to $0.83$, while the $5\sigma$ upper limits for non-detected sources span $f_{\mathrm{esc}}^{\mathrm{Ly}\alpha} < 0.03$ to $< 0.30$ (the full catalogue of individual measurements is given in Table~\ref{tab:sample_ids}). This scatter is clearly visible in Figure~\ref{fig:fesc_redshift} and is consistent with results from the broader literature at comparable redshifts \citep{Simmonds2023, Ning2023, Tang2023, Saxena2024, Lin2024}, where $f_{\mathrm{esc}}^{\mathrm{Ly}\alpha}$ spans more than an order of magnitude at fixed redshift. Taken together, these studies suggest that $f_{\mathrm{esc}}^{\mathrm{Ly}\alpha}$ rises with redshift up to $z \sim 5$--$6$, likely driven by declining dust content and increasingly porous ISM geometries \citep{Hayes2011, Konno2016, Lin2024}. Beyond $z \gtrsim 6.5$  however, the increasing neutral fraction of the IGM appears to become the dominant regulator of observed Ly$\alpha$ escape, suppressing the observed emission from individual galaxies \citep{Tang2023, Saxena2024, Jones2025}. Our sample lies close to this transition at $z \sim 6.1$, making it a useful benchmark for disentangling internal ISM and CGM effects from the onset of large-scale IGM attenuation.

As established in Section~\ref{sec:igm_uniformity}, the scatter observed within our sample is particularly notable given that all galaxies should experience broadly similar IGM conditions. The observed variation in $f_{\mathrm{esc}}^{\mathrm{Ly}\alpha}$ therefore likely reflects genuine galaxy-to-galaxy differences in ISM and CGM properties rather than large-scale environmental variations.

Given this large scatter, it is informative to consider a population-averaged escape fraction across the sample. Different approaches to averaging escape fraction datasets are used throughout the literature, so we report a suite of estimators in Table~\ref{tab:fesc_averages}. Considering only the Ly$\alpha$ detections gives a median escape fraction of $\langle f_{\mathrm{esc}}^{\mathrm{Ly}\alpha} \rangle_{\mathrm{det}} = 0.15$. This provides the most direct comparison to literature values derived only from Ly$\alpha$ detections, since such estimates are biased towards galaxies with favourable Ly$\alpha$ escape conditions regardless of how the parent sample was originally selected. Our sample, however, is H$\alpha$-selected and therefore complete with respect to star-formation rate down to the H$\alpha$ flux limit, rather than the underlying population as a whole. The non-detections therefore contain important information about the underlying galaxy population and should not be excluded. For illustration, we also report simple means treating upper limits either as their limit values or as zero (Table~\ref{tab:fesc_averages}), to show how different treatments of censored data affect the average. A relevant example from the recent literature is the $z \simeq 5.5$ H$\alpha$-selected sample of \citet{Cheng2026}, whose headline $\langle f_{\mathrm{esc}}^{\mathrm{Ly}\alpha} \rangle < 0.32$ is obtained by setting non-detections to their $3\sigma$ upper limits. This estimator is the analogue of our upper-limit-valued mean rather than of our adopted reverse Kaplan-Meier median, so the two should not be compared directly.

\begin{table}
\centering
\caption{Summary of population-averaged $f_{\mathrm{esc}}^{\mathrm{Ly}\alpha}$ estimators for the non-AGN JELS-MUSE sample ($N=22$; 10 detections and 12 upper limits) at $z \sim 6.1$. Different estimators are reported to facilitate comparison with literature studies adopting different treatments of censored data. The reverse Kaplan-Meier median and stacked flux estimate are adopted as our primary measurements, while the simple means illustrate the limiting cases for handling non-detections.}
\label{tab:fesc_averages}
\renewcommand{\arraystretch}{1.3}
\begin{tabular}{lc}
\hline
Estimator & $f_{\mathrm{esc}}^{\mathrm{Ly}\alpha}$ \\
\hline
Median (detections only) & $0.15$ \\
Mean (upper limits set to upper-limit values) & $0.18$ \\
Mean (upper limits set to zero) & $0.11$ \\
Stacked summed-flux estimate & $0.08^{+0.02}_{-0.02}$ \\
Reverse Kaplan--Meier median & $0.07^{+0.04}_{-0.03}$ \\
\hline
\end{tabular}
\end{table}

To incorporate the upper limits we apply reverse Kaplan-Meier survival analysis \citep{Feigelson1985, Jaskot2024a, Flury2025}. The standard Kaplan-Meier product-limit estimator \citep{Kaplan1958} was developed for right-censored data, where the true value lies above a known bound, as for a lifetime that exceeds the duration of a study. Ly$\alpha$ escape fractions derived from non-detections are instead left-censored, since each non-detection provides an upper limit. We therefore negate all values, which maps the left-censored upper limits onto the right-censored form the estimator expects, fit the Kaplan--Meier survival curve, identify the point where the survival function reaches 0.5, and negate the result to recover the median escape fraction. We implement this procedure using the \textsc{lifelines} package \citep{Davidson-Pilon2024}. Uncertainties are estimated through bootstrap resampling. We resample the full dataset with replacement 1000 times, recompute the reverse Kaplan--Meier median for each realisation, and adopt the 16th and 84th percentiles as the confidence interval. This yields $f_{\mathrm{esc}}^{\mathrm{Ly}\alpha} = 0.07^{+0.04}_{-0.03}$.

As an independent check we also calculate a stacked flux escape fraction. For each source we extract the continuum-subtracted 1D spectrum within the fixed $r=0.6$ arcsec aperture and integrate the flux in a window centred on the expected Ly$\alpha$ wavelength, using the fitted FWHM for detections and the median detected FWHM for non-detections. We sum the observed Ly$\alpha$ flux and the dust-corrected H$\alpha$ flux over all sources. We then convert these to luminosities as in Section~\ref{sec:haflux}. Non-detections therefore contribute their actual near-zero measured flux within this fixed window, rather than being estimated from pseudo-narrow-band imaging over the full JELS redshift range, which would be considerably noisier. The total Ly$\alpha$ luminosity is divided by $8.7 \times$ the summed dust-corrected H$\alpha$ luminosity, as in Equation~\ref{eq:fesc}. Bootstrap uncertainties are estimated by resampling sources with replacement 1000 times and, for each resampled source, drawing a perturbed flux from a Gaussian distribution centred on the measured value with standard deviation equal to the reported flux uncertainty, applied independently to the Ly$\alpha$ and H$\alpha$ measurements. This gives $f_{\mathrm{esc}}^{\mathrm{Ly}\alpha} = 0.08^{+0.02}_{-0.02}$, in excellent agreement with the reverse Kaplan-Meier estimate. We therefore adopt the reverse Kaplan-Meier median as our primary measurement.

Our measurements are in close agreement with the individual detections reported by \citet{Ning2023}, that have a median escape fraction of $0.07$. They are also broadly consistent with the H$\alpha$-anchored sample of \citet{Lin2024}, who stack Ly$\alpha$ in two redshift bins and find $f_{\mathrm{esc}}^{\mathrm{Ly}\alpha} = 0.086$ and $0.104$ at $z = 5.84$ and $5.24$. Although that study probes slightly lower redshifts, the H$\alpha$ sample is also emission-line selected, like JELS, making the comparison particularly meaningful. We note, however, that selections from slitless spectroscopic surveys such as FRESCO tend to be biased towards higher line flux and higher-equivalent-width emitters compared to narrow-band surveys \citep{Duncan2025}, and so selection effects are likely to impact this comparison. A closer redshift match is provided by \citet{Shimizu2026}, who select H$\alpha$ emitters at $z \simeq 6.2$ directly via dual narrow-band imaging, combining \textit{JWST}/NIRCam F470N with Subaru/HSC NB872, rather than slitless spectroscopy. Stacking their sample, they obtain a completeness-weighted median $f_{\mathrm{esc}}^{\mathrm{Ly}\alpha} = 0.106^{+0.066}_{-0.044}$ ($0.090^{+0.063}_{-0.032}$ unweighted), consistent with our own reverse Kaplan-Meier and stacked-flux estimates within the combined uncertainties. Together, these studies support a picture in which the typical population-averaged Ly$\alpha$ escape fraction at $z \sim 5$--$6$ is relatively modest once non-detections and selection effects are properly accounted for.

However, our average values do lie below several luminosity-function-based estimates and individual Ly$\alpha$-bright samples at comparable redshifts, including the relations presented by \citet{Hayes2011} and \citet{Konno2016}, as well as the individual detections reported by \citet{Saxena2024} and \citet{Simmonds2023}, as shown in Figure~\ref{fig:fesc_redshift}. Part of this offset likely reflects methodological differences. Luminosity-function based estimates depend on the adopted integration limits, faint-end slopes, and UV to H$\alpha$ star-formation rate calibrations, all of which can introduce systematic offsets in the inferred $f_{\mathrm{esc}}^{\mathrm{Ly}\alpha}$ at fixed redshift \citep{Konno2016, Sun2023}. As discussed above, the comparison with Ly$\alpha$-selected samples, which may be biased towards higher values, likely also contributes to this offset. The relatively low values we measure are also consistent with the onset of mild IGM attenuation at $z \sim 6.1$. As the transmission distributions in Section~\ref{sec:igm_uniformity} show (Figure~\ref{fig:igm_transmission}), some IGM suppression is already present at this redshift, which will act to lower the population-averaged Ly$\alpha$ signal and can therefore account for part of the offset we see. This is consistent with the gradual decline in average Ly$\alpha$ visibility inferred beyond $z \gtrsim 6$ from both spectroscopic surveys and luminosity-function studies. 

To consider this measurement in the wider context of reionization, we can use the empirical $f_{\mathrm{esc}}^{\mathrm{LyC}} \simeq 0.15^{+0.06}_{-0.04} \, f_{\mathrm{esc}}^{\mathrm{Ly}\alpha}$ relation found by \citet{Begley2024} for $z \simeq 4$--$5$ star-forming galaxies to extrapolate our population-averaged $f_{\mathrm{esc}}^{\mathrm{Ly}\alpha} = 0.07^{+0.04}_{-0.03}$ to an approximate LyC escape fraction. This gives $f_{\mathrm{esc}}^{\mathrm{LyC}} \approx 0.15 \times 0.07 \approx 0.01$, of order 1 per cent. Applying this relation to our sample requires some assumptions. The \citet{Begley2024} calibration is derived at $z \simeq 4$--$5$ from a UV-continuum-selected VANDELS sample, using composite-spectrum low-ionization state (LIS) absorption strength as an indirect proxy for $f_{\mathrm{esc}}^{\mathrm{LyC}}$, whereas our galaxies are H$\alpha$-selected at $z \sim 6.1$, moving into the EoR. Applying their relation to our sample therefore assumes that the physical link between Ly$\alpha$ and LyC escape, and its normalisation, carries over both to a different selection function and to somewhat higher redshift, neither of which is guaranteed given the evolving ISM conditions expected across this interval. It also assumes that a relation calibrated on individual galaxies and composite spectra applies equally well to a population-averaged $f_{\mathrm{esc}}^{\mathrm{Ly}\alpha}$ derived through survival analysis, which combines detections and non-detections in a different way to the object-by-object or fixed-EW stacking used in their study. With these caveats in mind, we regard this LyC estimate as an illustrative extrapolation rather than a direct measurement for our sample.

This average-based approach may not, however, be the most informative way to frame the ionizing contribution of our sample. The large scatter in $f_{\mathrm{esc}}^{\mathrm{Ly}\alpha}$ has implications beyond the population average itself. \citet{Giovinazzo2026} find that strongly leaking galaxies ($f_{\mathrm{esc}} > 10$ per cent), only around 20 per cent of their sample, produce roughly 87 per cent of the total escaping ionizing output, while the far more numerous weak leakers contribute negligibly. A low population-averaged escape fraction is therefore not inconsistent with reionization being driven by a small, highly leaking subset of galaxies. Given the comparably wide spread we find, from $0.04$ to $0.83$ among individual detections, the averaged values reported above should not be read as representative of every galaxy's contribution to reionization. Instead, a broader understanding of ionizing escape requires us to consider the full distribution of $f_{\mathrm{esc}}^{\mathrm{Ly}\alpha}$, including its upper tail, where the more highly leaking subset could lie. In Section~\ref{sec:proprelations} we explore how this picture, in which the galaxies that dominate the ionizing budget are outliers within the population rather than typical members of it, could indicate episodic Ly$\alpha$ and LyC escape.

\subsection{Ly{\fontsize{10}{10}\selectfont$\mathbf{\alpha}$} Escape Fraction as a Function of Galaxy Properties}
\label{sec:proprelations}

Observable galaxy properties provide indirect probes of the ISM conditions governing Ly$\alpha$ radiative transfer. Because Ly$\alpha$ photons resonantly scatter in neutral hydrogen, their escape depends sensitively on the neutral gas column density, covering fraction, velocity structure, and dust geometry within the ISM and CGM \citep{Dijkstra2014, Verhamme2015, Verhamme2017}. This sample is particularly well suited to examining how $f_{\mathrm{esc}}^{\mathrm{Ly}\alpha}$ varies across the galaxy population because it is H$\alpha$ selected rather than selected on Ly$\alpha$ emission itself. It therefore spans the full range of Ly$\alpha$ visibility, including systems where Ly$\alpha$ is weak or entirely undetected, reducing the selection bias inherent to Ly$\alpha$-selected samples. As shown in Section~\ref{sec:igm_uniformity}, the galaxies are also expected to experience broadly similar IGM transmission, so variations in $f_{\mathrm{esc}}^{\mathrm{Ly}\alpha}$ can be interpreted primarily in terms of intrinsic galaxy properties rather than large fluctuations in IGM opacity.

\begin{figure*}
	\includegraphics[width=\textwidth]{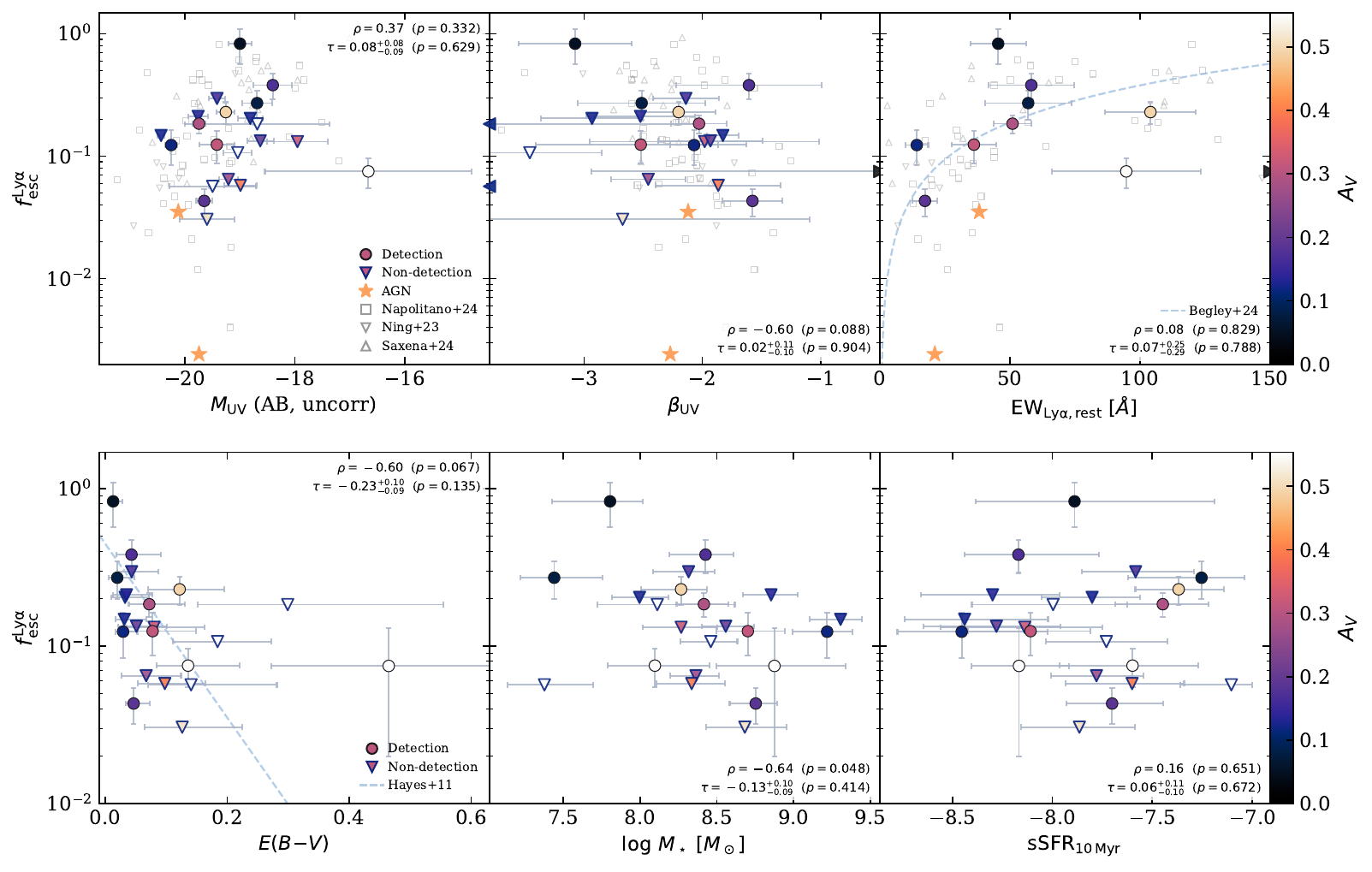}
	\caption{Ly$\alpha$ escape fraction as a function of (Left to right) UV absolute magnitude ($M_{\mathrm{UV}}$), UV spectral slope ($\beta$), Ly$\alpha$ rest-frame equivalent width ($\mathrm{EW}_{\mathrm{Ly}\alpha,\mathrm{rest}}$), nebular dust extinction  ($E(B-V)$), stellar mass ($\mathrm{M}_{\star}$), and specific star-formation rate ($\mathrm{sSFR}_{10\mathrm{Myr}}$) for the H$\alpha$-selected galaxy sample at $z \simeq 6.1$. Filled circles show Ly$\alpha$ detections and triangles indicate $5\sigma$ upper limits for Ly$\alpha$ non-detections, both coloured by the SED-derived dust attenuation ($A_V$). The two AGN are shown in the upper panels as stars.  Sources with $\beta$ values outside the range expected for physically plausible stellar populations are shown with horizontal arrows at the panel boundaries. The dashed line in the $E(B-V)$ panel shows the empirical attenuation relation from \citet{Hayes2011}. Literature comparison samples from \citet{Ning2023}, \citet{Saxena2024}, and \citet{Napolitano2024} are plotted where available. Both the Spearman rank correlation coefficient using the Ly$\alpha$ detections alone and the censored Kendall $\tau$ statistic including upper limits are shown for each plot. }
	\label{fig:esc_prop}
\end{figure*}

Figure~\ref{fig:esc_prop} shows $f_{\mathrm{esc}}^{\mathrm{Ly}\alpha}$ as a function of UV absolute magnitude ($M_{\mathrm{UV}}$), UV spectral slope ($\beta$), Ly$\alpha$ rest-frame equivalent width $\mathrm{EW}_{\mathrm{Ly}\alpha,\mathrm{rest}}$, nebular dust extinction ($E(B-V)$; see Section~\ref{sec:sed}), stellar mass ($M_\star$), and specific star-formation rate (sSFR). These corresponding physical properties for each source are tabulated in Table~\ref{tab:sample_props}. To assess possible trends, we compute both the Spearman rank correlation coefficient using the Ly$\alpha$ detections alone and the censored Kendall $\tau$ statistic including upper limits. The latter follows the formalism of \citet{Akritas1996}, implemented following \citet{Flury2023kendall}, allowing the non-detections to contribute without discarding censored measurements. The resulting statistics are reported in Table~\ref{tab:correlations}.

\begin{table}
\centering
\caption{Correlation statistics between $f_{\mathrm{esc}}^{\mathrm{Ly}\alpha}$ and galaxy properties shown in Figure~\ref{fig:esc_prop}. Spearman rank coefficients are computed using Ly$\alpha$ detections only, while the censored Kendall $\tau$ statistic includes upper limits following the formalism of \citet{Akritas1996} implemented via \citet{Flury2023kendall}. Uncertainties on the Kendall $\tau$ values are derived through bootstrap resampling.}
\label{tab:correlations}
\renewcommand{\arraystretch}{1.4}
\begin{tabular}{lcccc}
\hline
Property & $\rho_{\mathrm{Spearman}}$ & $p_{\mathrm{Spearman}}$ & $\tau_{\mathrm{Kendall}}$ & $p_{\mathrm{Kendall}}$ \\
\hline
$M_{\mathrm{UV}}$ & 0.37 & 0.332 & $0.08^{+0.08}_{-0.09}$ & 0.629 \\
$\beta$ & -0.60 & 0.088 & $0.02^{+0.11}_{-0.10}$ & 0.904 \\
$\mathrm{EW}_{\mathrm{Ly}\alpha,\mathrm{rest}}$ & 0.08 & 0.829 & $0.07^{+0.25}_{-0.29}$ & 0.788 \\
$E(B-V)$ & -0.60 & 0.067 & $-0.23^{+0.10}_{-0.09}$ & 0.135 \\
$\log(M_\star/\mathrm{M}_\odot)$ & -0.64 & 0.048 & $-0.13^{+0.10}_{-0.09}$ & 0.414 \\
$\log(\mathrm{sSFR}_{10\mathrm{Myr}}/\mathrm{yr}^{-1})$ & 0.16 & 0.651 & $0.06^{+0.11}_{-0.10}$ & 0.672 \\
\hline
\end{tabular}
\end{table}

Although the trends discussed below are not all individually significant at high formal confidence, several point consistently in the same direction across both the Spearman and censored Kendall statistics, and are worth examining qualitatively. This is expected given the sample size of 10 Ly$\alpha$ detections, after excluding the AGN, and 12 upper limits, and that previous studies have consistently found large intrinsic scatter in $f_{\mathrm{esc}}^{\mathrm{Ly}\alpha}$ at fixed galaxy property across a wide range of redshifts \citep{Hayes2011, Gronke2016, Verhamme2017, Tang2024, Napolitano2024, Lin2024}. With a limited sample size and dynamic range, strong formal statistical significance is difficult to recover even where genuine physical trends are present. We therefore focus primarily on the direction and consistency of the observed relations.

The clearest tendencies are seen with nebular dust extinction, UV slope, and stellar mass, in the sense that galaxies with larger Ly$\alpha$ escape fractions tend to be less dusty, bluer, and lower mass. The detections broadly follow the attenuation relation of \citet{Hayes2011}. Similar behaviour has been reported with far greater statistical power in the larger H$\alpha$-selected samples of \citet{Lin2024} and \citet{Napolitano2024}, and in the dual-narrow-band sample of \citet{Shimizu2026}, all of which find the strongest correlations with dust-sensitive quantities such as $E(B-V)$ and $\beta$, alongside weaker trends with stellar mass and little dependence on UV luminosity, qualitatively matching the behaviour seen here. Notably, \citet{Begley2022} find the same qualitative trends in their constraints on $f_{\mathrm{esc}}^{\mathrm{LyC}}$, suggesting these dependencies may reflect a common physical origin across both Ly$\alpha$ and LyC escape.

\emph{$E(B-V)$ and $\beta$:} We find anti-correlations between $f_{\mathrm{esc}}^{\mathrm{Ly}\alpha}$ and both $E(B-V)$ and $\beta$. The two are comparable on the detections alone, but the trend with $E(B-V)$ is marginally more robust once upper limits are included through the censored Kendall $\tau$. This difference is not significant and may simply reflect the poorer constraints on $\beta$ for some sources rather than a genuine physical distinction between the two tracers. Similar anti-correlations with both quantities were reported by \citet{Lin2024} for a larger H$\alpha$-selected sample at comparable redshift. These dust-sensitive trends are consistent with feedback-regulated escape scenarios, in which young stellar populations photoionize and disrupt their surrounding gas clouds, temporarily creating lower-column-density channels through which Ly$\alpha$ photons can preferentially escape \citep{Verhamme2015, Jaskot2019, Flury2022a}. In this picture Ly$\alpha$ escape is governed less by the total dust content of the galaxy, and more by the local neutral gas geometry and optical depth along particular sightlines. This is consistent with radiative transfer models in which escape depends on the covering fraction, column density, clumpiness, and outflow structure of the neutral gas rather than dust content alone \citep{Verhamme2015, Gronke2016, Verhamme2017}, and with UV absorption-line studies of low-redshift Lyman continuum emitters that find similarly clumpy, non-uniform neutral gas \citep{Gazagnes2018, Gazagnes2020, Saldana-Lopez2022, Mauerhofer2021, Flury2025, Mauerhofer2025}. We note that dust and neutral gas geometry are difficult to separate observationally, since the reddening and UV slope tend to decline together with the covering fraction and optical depth of the neutral gas \citep{Flury2025}, so the dust-sensitive trends seen here may partly reflect this shared dependence. Because $f_{\mathrm{esc}}^{\mathrm{Ly}\alpha}$ is computed from dust-corrected H$\alpha$ and so carries an implicit dependence on $E(B-V)$, we recompute it without the dust correction and find that all of the property correlations shift rather than the $E(B-V)$ trend alone, confirming that this dependence does not by itself account for the observed anti-correlation.

\emph{Stellar mass and sSFR:} The stellar mass trend may reflect the shallower potentials of low-mass galaxies, which allow stellar feedback to clear neutral gas more easily, whereas the deeper potentials of more massive systems can suppress feedback and inhibit the clearing of escape channels. On this basis dwarf galaxies have long been favoured as Ly$\alpha$ and LyC emitters \citep{Razoumov2010, Wise2014, Paardekooper2015}. Our anti-correlation with stellar mass matches that of \citet{Napolitano2024}, interpreted there as reflecting the greater neutral gas and dust content of more massive systems. We recover no clear trend with sSFR measured over a $10$ Myr timescale, the interval most closely matched to the feedback that regulates Ly$\alpha$ escape, although this is also the noisiest quantity to constrain from the SED fits. The $100$ Myr sSFR shows a nominally stronger but still insignificant correlation, so we do not find the tighter link with the more recent star formation that a feedback-driven interpretation would predict. Such a picture, in which elevated sSFR marks systems in intense recent star formation, would have feedback lower the neutral gas covering fraction and open transient low-column-density channels \citep{Cen2015, Dijkstra2016, Sharma2017, Kimm2019, Maji2022}. The present uncertainties are too large to test this, and the considerable scatter of $f_{\mathrm{esc}}^{\mathrm{Ly}\alpha}$ against sSFR seen at low redshift suggests a larger sample will be needed to recover any underlying trend.

\emph{$M_{\mathrm{UV}}$ and Ly$\alpha$ equivalent width:} We find no strong correlation between $f_{\mathrm{esc}}^{\mathrm{Ly}\alpha}$ and either $M_{\mathrm{UV}}$ or $\mathrm{EW}_{\mathrm{Ly}\alpha,\mathrm{rest}}$. Similar weak or absent trends with UV luminosity have been reported previously \citep{Hayes2011, Verhamme2017, Lin2024}. We do, however, note a weak qualitative tendency for the fraction of strong leakers ($f_{\mathrm{esc}}^{\mathrm{Ly}\alpha} > 10$ per cent) to increase towards UV-fainter magnitudes, which the rank correlation coefficient does not capture. This would be consistent with scenarios in which faint galaxies are significant contributors to the reionization photon budget \citep{Mascia2024, Jecmen2026}. Our sample is too small to quantify this split reliably, so we report it only as a qualitative feature.  We do not recover a significant $f_{\mathrm{esc}}^{\mathrm{Ly}\alpha}$--$\mathrm{EW}$ relation, though our detections remain consistent with the positive trends found in samples spanning a wider EW range \citep{Ning2023, Begley2024, Napolitano2024, Saxena2024, Shimizu2026}. The absence of a strong correlation here is therefore most likely driven by our narrow dynamic range in $\mathrm{EW}_{\mathrm{Ly}\alpha,\mathrm{rest}}$, combined with the modest sample size and substantial intrinsic scatter in Ly$\alpha$ escape.

A central result of this analysis is therefore that no single galaxy property predicts $f_{\mathrm{esc}}^{\mathrm{Ly}\alpha}$, with galaxies of similar attenuation, mass or UV slope spanning more than an order of magnitude in escape fraction. This points to Ly$\alpha$ escape being governed by the small-scale structure of the neutral gas along the line-of-sight, which the integrated properties measured here only weakly trace. Integrated Ly$\alpha$ fluxes alone cannot recover this structure either, since the degeneracies between covering fraction, gas kinematics, dust distribution, and \ion{H}{i} column density limit how much can be extracted from the escape fraction alone \citep{Verhamme2015, Verhamme2017}, so similar escape fractions can arise from quite different neutral gas configurations.

As established in Section~\ref{sec:igm_uniformity}, the scatter most likely reflects intrinsic diversity in ISM and CGM structure near the end of reionization rather than variations in IGM opacity. Some of this scatter may also reflect the viewing angle of individual sightlines through an anisotropic, picket-fence-like ISM, since Ly$\alpha$ escape can vary substantially depending on whether a given line-of-sight intersects an ionized, low-column-density channel \citep{Heckman2001, Behrens2014, Gazagnes2018}. This is consistent with radiative transfer models in which escape depends strongly on the covering fraction, column density, clumpiness, and outflow structure of the neutral gas rather than dust content alone \citep{Gronke2016, Verhamme2017}.

Temporal variability may also contribute significantly to the observed scatter. Feedback from young stellar populations operates across multiple timescales and so photoionization and radiation pressure can begin clearing their surrounding gas clouds within a few Myr, while supernova-driven outflows emerge over longer timescales of $\sim10$--40 Myr \citep{Jaskot2019, Flury2022a}. Supernovae are far more effective at clearing gas than radiation pressure or photoionization, but producing them requires the death of the same massive stars that generate the Ly$\alpha$ and LyC photons \citep{Jaskot2025}. Bursty star formation on $\sim10$ Myr timescales offers a natural resolution, since supernovae from one generation of stars can clear low-column-density channels through which the ionizing photons of a subsequent generation then escape \citep{Jaskot2019, Flury2022a, Jaskot2025, Flury2025, LeReste2025}. If star-formation histories are bursty, galaxies with otherwise similar global properties may therefore be observed at very different stages of their feedback cycle. In this scenario, Ly$\alpha$ escape may be highly time-variable, with transient phases of enhanced escape occurring when feedback temporarily reduces the local neutral gas column density or covering fraction. The small subset of strong leakers that dominates the ionizing output in Section~\ref{sec:fesc_evolution} \citep{Giovinazzo2026} may then be galaxies caught during such a phase, rather than a permanently distinct population, so that the strongest Ly$\alpha$ and LyC leakers seen at any one epoch are a rotating minority drawn from the wider population. Episodic escape of this kind was inferred for the $z \simeq 0.3$ Lyman continuum emitters of \citet{Flury2022b} on similar statistical grounds, and is supported by subsequent analysis of their stellar populations \citep{Flury2025} and radio continuum shapes \citep{Bait2024}. Such a cycle would help account for the large scatter in $f_{\mathrm{esc}}^{\mathrm{Ly}\alpha}$ we measure at fixed galaxy property.

In future work we will use Ly$\alpha$ line-profile modelling with \textit{JWST}/NIRSpec data from the DJA \citep{deGraaff2024, Heintz2024} to provide systemic redshifts and probe these conditions more directly \citep{Gronke2015, Gronke2017}. While the present sample already represents the majority of the H$\alpha$-selected galaxy population within this cosmic volume, the completion of the full JELS-MUSE survey and its combination with larger H$\alpha$-selected samples from programmes such as \textit{JWST} Cycle 4 MINERVA programme (GO 7814; PI: A. Muzzin) and \textit{JWST} Cycle 3 COSMOS-3D programme (GO  5893; PI: K. Kakiichi) will expand the dynamic range and cosmic volume available for comparison. This will improve the statistical power of correlation analyses and allow a more detailed investigation of how Ly$\alpha$ escape correlates with galaxy properties, ISM structure, feedback, and radiative transfer conditions near the end of reionization.


\section{Summary}
\label{sec:summary}

In this paper, we have presented a study of Ly$\alpha$ escape fractions for a complete, H$\alpha$-flux-limited sample of star-forming galaxies at $z \approx 6.1$, selected from the \textit{JWST} Emission Line Survey \citep[JELS;][]{Duncan2025, Pirie2025} and observed in Ly$\alpha$ with VLT/MUSE as part of the JELS-MUSE Large Area Survey (Li et al., in prep.). By anchoring the Ly$\alpha$ detections to a complete sample of H$\alpha$ emitters in a contiguous field at a narrow redshift slice, we substantially reduce the Ly$\alpha$ and UV selection biases that affect samples selected on those properties directly, providing a more representative census of $f_{\mathrm{esc}}^{\mathrm{Ly}\alpha}$ across the star-forming galaxy population at this epoch. Our main conclusions are summarised here.

\begin{itemize}

\item \textbf{Detection fraction.} We detect Ly$\alpha$ emission in 12 of 24 H$\alpha$-selected sources within the current MUSE footprint, corresponding to a Ly$\alpha$ detection fraction of $50 \pm 10$ per cent, where the uncertainties are binomial. Of these 12 detections, 7 exceed our strong-detection threshold $\Sigma^{99.5}_{\mathrm{Ly}\alpha}$, while the 12 sources below the detection threshold $\Sigma^{98}_{\mathrm{Ly}\alpha}$ are classified as non-detections and adopted as upper limits in the analysis. Both thresholds are defined empirically from the false-positive analysis of Section~\ref{sec:lyadetectsig}. To compare like-for-like with the UV-faint Lyman-break samples of \citet{Stark2011} at $z \sim 6$, we restrict to their fiducial magnitude range $-20.25 < M_{\mathrm{UV}} < -18.75$ and account for sources where the local sensitivity could not have recovered a canonical emitter, giving a Ly$\alpha$ emitter fraction of $X_{\mathrm{Ly}\alpha} = 33 \pm 12$ per cent (5 of 15) for $\mathrm{EW}(\mathrm{Ly}\alpha) > 25$ \AA. This is consistent with the $54 \pm 11$ per cent of \citet{Stark2011} within the binomial uncertainties, and lies fully within the field-to-field range predicted by inhomogeneous reionization models \citep{Napolitano2024}.

\item \textbf{Drivers of Ly$\alpha$ detectability.} The non-detections are not preferentially faint in the UV ($D_{\mathrm{KS}} = 0.25$, $p_{\mathrm{KS}} = 0.85$), confirming that the absence of Ly$\alpha$ reflects genuine variation in ISM escape conditions rather than insufficient sensitivity. We find no statistically significant difference between Ly$\alpha$-detected and undetected sources in $A_V$ ($p_{\mathrm{KS}} = 0.63$), stellar mass ($p_{\mathrm{KS}} = 0.80$), or $\mathrm{SFR}_{10\,\mathrm{Myr}}$ ($p_{\mathrm{KS}} = 0.56$). This is consistent with Ly$\alpha$ escape being governed by neutral gas covering fraction and column density on sub-kiloparsec scales that are not captured by broadband photometric diagnostics.

\item \textbf{Ly$\alpha$ escape fraction distribution.} The detected sources span a wide range of $f_{\mathrm{esc}}^{\mathrm{Ly}\alpha} = 0.04$--$0.83$, with $5\sigma$ upper limits for non-detected sources spanning $f_{\mathrm{esc}}^{\mathrm{Ly}\alpha} < 0.03$ to $< 0.30$. For the non-AGN sample of 22 sources (10 detections and 12 upper limits), incorporating the non-detections as upper limits through reverse Kaplan-Meier survival analysis yields a population-averaged escape fraction of $f_{\mathrm{esc}}^{\mathrm{Ly}\alpha} = 0.07^{+0.04}_{-0.03}$, in excellent agreement with an independent stacked-flux estimate of $0.08^{+0.02}_{-0.02}$. Because the sample is H$\alpha$-selected and complete above the H$\alpha$ flux limit, this measurement does not require corrections for faint-end slopes or UV luminosity-function uncertainties, making it directly comparable to the emerging class of \textit{JWST}-enabled H$\alpha$-anchored measurements at $z \approx 5.5$--$6.5$ \citep{Lin2024, Shimizu2026}.

\item \textbf{Redshift evolution.} Our population-averaged $f_{\mathrm{esc}}^{\mathrm{Ly}\alpha}$ at $z \approx 6.1$ is consistent with the direct \textit{JWST}-based determinations from the H$\alpha$-anchored studies of \citet{Lin2024} at $z \approx 4.9$--$6.3$ and \citet{Shimizu2026} at $z \simeq 6.2$, and with the individual detections of \citet{Ning2023}, but lies below several luminosity-function based estimates and Ly$\alpha$-selected samples at comparable redshifts \citep{Hayes2011, Konno2016, Simmonds2023, Saxena2024}. This offset likely reflects the combined effect of selection bias in Ly$\alpha$-selected samples and systematic uncertainties in luminosity-function integration, and may additionally indicate the onset of mild IGM attenuation at $z \approx 6.1$. The large scatter in $f_{\mathrm{esc}}^{\mathrm{Ly}\alpha}$ within our single-field, narrow-redshift sample, where simulations indicate the IGM transmission is expected to be broadly uniform (Section~\ref{sec:igm_uniformity}), shows that substantial galaxy-to-galaxy intrinsic variance is present at this epoch and must be accounted for when interpreting the declining $f_{\mathrm{esc}}^{\mathrm{Ly}\alpha}$ observed at $z \gtrsim 6.5$ \citep{Saxena2024, Tang2024}, where the neutral IGM becomes the dominant regulator of Ly$\alpha$ visibility.

\item \textbf{Dependence on galaxy properties.} The clearest tendencies in $f_{\mathrm{esc}}^{\mathrm{Ly}\alpha}$ are with nebular dust extinction $E(B-V)$, UV spectral slope $\beta$, and stellar mass $M_\star$, in the sense that galaxies with larger escape fractions tend to be less dusty, bluer, and lower mass (Spearman $\rho = -0.60$, $-0.60$, and $-0.64$ for the detections). Once the upper limits are included through the censored Kendall $\tau$ statistic, only the trend with $E(B-V)$ retains a comparable strength ($\tau = -0.23^{+0.10}_{-0.09}$), while the $\beta$ and $M_\star$ trends weaken. None of the correlations are individually significant at high confidence, so we do not draw quantitative constraints from them, but their direction is qualitatively consistent with previous studies at both low and high redshift \citep{Verhamme2017, Flury2022b, Lin2024, Napolitano2024} and points to a picture in which Ly$\alpha$ escape is regulated by feedback-driven low-column-density channels in the ISM surrounding young stellar populations. Crucially, galaxies with similar $E(B-V)$, mass or UV slope span more than an order of magnitude in $f_{\mathrm{esc}}^{\mathrm{Ly}\alpha}$, demonstrating that no single galaxy property predicts Ly$\alpha$ escape for individual systems and underscoring the dominant role of small-scale neutral gas geometry and ISM structure.

\end{itemize}

Our sample is drawn from a single contiguous field within a narrow redshift interval ($\Delta z \approx 0.1$), and simulations indicate the sources are expected to experience broadly similar IGM transmission near the end of reionization (Section~\ref{sec:igm_uniformity}). The observed scatter in $f_{\mathrm{esc}}^{\mathrm{Ly}\alpha}$ therefore provides an empirical baseline for the intrinsic variance at this epoch, against which the additional suppression by the rising neutral IGM at $z \gtrsim 6.5$ can be isolated. As the JELS-MUSE Large Area Survey progresses to its full 77-pointing mosaic and deeper NIRSpec spectroscopy provides direct H$\alpha$ measurements for individual sources, the constraints presented here will be substantially sharpened, establishing this field as a benchmark for intrinsic Ly$\alpha$ escape physics at the end of the Epoch of Reionization. 


\section*{Acknowledgements}

ALP is supported by a Science and Technology Facilities Council (STFC) PhD studentship. KJD acknowledges support from STFC through an Ernest Rutherford Fellowship (grant number ST/W003120/1), and CAP acknowledges STFC support via grants ST/W507441/1 and ST/Y000951/1. ZL and AMS acknowledge STFC grant ST/X001075/1. LK acknowledges the support of a Royal Society University Research Fellowship (grant number URF\textbackslash R1\textbackslash251793). SRF acknowledges support from Leverhulme ECF-2025-361. DJM and JSD acknowledge the support of the Royal Society through the award of  a Royal Society Research Professorship to JSD. E.I. gratefully acknowledge financial support from ANID - MILENIO - NCN2024\_112 and ANID FONDECYT Regular 1221846. Some of the data products presented herein were retrieved from the Dawn \textit{JWST} Archive (DJA), an initiative of the Cosmic Dawn Center (DAWN), which is funded by the Danish National Research Foundation under grant DNRF140.

\section*{Data Availability}

The JELS data underlying this article are available in the Mikulski Archive for Space Telescopes (MAST) at \url{https://doi.org/10.17909/8v6n-ad45}. Higher level data products, including reduced mosaics in the JELS filters and other \textit{JWST} and \textit{HST} broad-band filters, as well as associated catalogues is publicly available through the University of Edinburgh \href{https://datashare.ed.ac.uk}{DataShare}. The JELS-MUSE observations were obtained under ESO programme [112.25WM.003], and the raw data are publicly available through the ESO Science Archive. A survey paper presenting the fully reduced datacubes and corresponding source catalogues will be released on completion of the full survey in 2027 (Li et al. in prep.). Any other data produced for this article will be shared on reasonable request to the corresponding author. For the purpose of open access, the author has applied a Creative Commons Attribution (CC BY) licence to any Author Accepted Manuscript version arising from this submission.



\bibliographystyle{mnras}
\bibliography{ref} 




\appendix

\section{Astrometric Corrections and Mosaic Construction}
\label{appendix:astrometry}
Astrometric alignment was performed relative to \textit{HST}/ACS F814W imaging in the COSMOS field \citep{Koekemoer2007, Scoville2007}. Both this reference imaging and the \textit{JWST} data are registered to the \textit{Gaia} astrometric frame, so the choice between them is not one of absolute accuracy but of which provides the cleaner cross-match. We align to the F814W imaging because its passband overlaps the MUSE wavelength range, so the same continuum sources are detected with comparable morphology in both frames, making it a more robust reference for centroid matching than the near-infrared \textit{JWST} imaging. Corrections were derived using white-light images generated from each individual MUSE exposure.

The primary alignment method follows a multi-source comparison approach similar to that adopted in previous deep MUSE surveys \citep{Bacon2017}. For each exposure, untargeted source catalogues were generated from both the MUSE white-light image and a PSF-matched \textit{HST} cutout using a \texttt{photutils} implementation of \textsc{SExtractor} \citep{Bertin1996}. The \textit{HST}/ACS F814W image was convolved to match the approximate MUSE white-light PSF (FWHM $\approx 0.7$ arcsec) prior to source detection. Sources were filtered to retain compact, isolated objects by applying radial and nearest-neighbour cuts. Edge sources and objects too faint to be reliably detected in the MUSE data were excluded. The catalogues were cross-matched using a 2 arcsec radius to ensure recovery of sources in exposures with large initial offsets.

For each matched source, the RA and Dec offsets were computed. A $3\sigma$ clipping procedure was applied to remove outliers, and the median RA and Dec shift was adopted as the global astrometric correction. This shift was applied to the MUSE WCS header. The automated procedure aligned the majority of exposures to sub-pixel accuracy, with residual offsets $\lesssim 0.2$ arcsec after correction. These residuals were verified by re-cross-matching each corrected exposure to the \textit{HST} reference and measuring the offset that remained. Exposures whose post-correction residual exceeded the $0.2$ arcsec MUSE pixel scale were flagged for manual correction. We also confirmed that any rotational offset between the MUSE and \textit{HST} frames was negligible, so only translational shifts in RA and Dec were applied. This was checked by fitting a similarity transform, allowing for a rotation, a uniform scale, and a translation, between the matched MUSE and \textit{HST} source positions, and reading the rotation angle from the fit. The recovered rotation was $ 0.2$ degrees, which displaces a source at the corner of the $1\ \mathrm{arcmin}$ field of view by $\lesssim 0.15$ arcsec, below the $0.2$ arcsec spatial sampling.

Automatic alignment failed for 15 per cent of the individual exposures and so these were corrected manually. The primary reason for failure is when the initial WCS offset is large enough to approach or exceed the $2$ arcsec cross-match radius, so that genuine counterparts fall outside the match window and too few reliable source pairs survive for the sigma-clipped median shift to be well determined. Increasing the radius, however, increases false matches reducing the overall efficacy of the method. This can also be compounded by a small number of usable sources in the affected exposures, either because they lie towards the edge of the mosaic or because their depth or seeing leaves few compact objects detectable in both frames. The manual corrections were measured by displaying the original MUSE white-light image and the non-convolved \textit{HST} cutout in \texttt{ds9} \citep{SmithsonianAstrophysicalObservatory2000} with WCS coordinates locked. Offsets were derived by interactively selecting compact sources in both images and iteratively adjusting the RA and Dec shifts until visual alignment was achieved, again reaching sub-pixel residuals ($\lesssim 0.2$ arcsec). After this step, we have the offsets required to align each cube and these can be applied and the exposures mosaicked into one cube.

\begin{figure}
\centering
\includegraphics[width=\columnwidth]{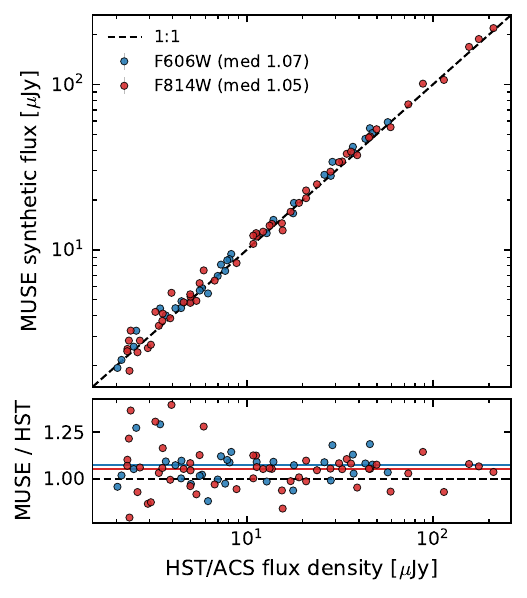}
\caption{Flux calibration check on the mega-cube in the \textit{HST}/ACS F606W (blue) and F814W (red) bands. For each compact star ($\mathrm{CLASS\_STAR} > 0.9$) within the mosaic footprint, a synthetic flux density is built by folding its MUSE spectrum, extracted in a $2$ arcsec aperture, through the relevant ACS/WFC throughput, then compared to the catalogue flux density in the same band. The upper panel shows the one-to-one comparison, with the dashed line marking equality. The lower panel shows the MUSE-to-\textit{HST} ratio against \textit{HST} flux density, with the median for each band drawn as a solid line and unity as the dashed line. The median ratios are $1.07$ in F606W and $1.05$ in F814W.}
\label{fig:flux_cal}
\end{figure}
Prior to mosaicking, all cubes were trimmed to a common wavelength grid to ensure slice-by-slice consistency. The final grid spans 4749.9--9349.9\,\AA\ with 1.25\,\AA\ sampling (3681 spectral slices), corresponding to the minimum common wavelength coverage across exposures. A narrow spectral region around the sodium D wavelength ($\sim 5890$\,\AA) was masked due to contamination from the sodium laser guide star used for adaptive optics correction.

The final ``mega-cube'' was constructed by stacking the aligned mosaic slices, with the wavelength assigned to each slice corresponding to the median wavelength of contributing exposures. The primary header records the applied offsets and contributing exposures. The variance was propagated through the same slice-by-slice construction. For each wavelength slice the reprojected \texttt{STAT} extensions of the contributing exposures were combined by inverse-variance summation, so that the combined variance is $\left(\sum_i 1/\sigma_i^2\right)^{-1}$, with non-finite or non-positive pixels excluded from the sum. Stacking these per-slice variance mosaics produces the variance cube, which is attached as the \texttt{STAT} extension of the flux mega-cube. We compared this propagated per-pixel noise against empirical noise estimates measured directly from wavelength slices across the full spectral range. The propagated variance was typically a factor of two higher than the empirical estimate, likely because the latter smooths over small-scale variation, so we retain the propagated variance as the more conservative estimate. The final data product is compatible with standard \texttt{MUSE MPDAF} \citep{Bacon2016} analysis tools.

To verify that the flux scale is preserved through the slice-by-slice combination, we compared synthetic photometry measured from the mega-cube against \textit{HST}/ACS catalogue fluxes for compact point sources, in both the F606W and F814W bands. As above, F814W is the reference band used for the astrometric alignment and so the F606W provides an additional independent check. Stars were selected from the \textit{HST} catalogue as objects with $\mathrm{CLASS\_STAR} > 0.9$ and positive flux, restricted to those lying within the mosaic footprint. The F606W sample spans flux densities of roughly $2$ to $60\,\mu$Jy and the F814W sample roughly $2$ to $200\,\mu$Jy. For each source, the spectrum was extracted across the relevant bandpass from the mega-cube in a circular aperture of $2$ arcsec radius, with a local background subtracted per wavelength slice from a surrounding annulus. The $2$ arcsec aperture is adopted as the total flux, justified by the curve of growth, which plateaus by $\approx 1.5$ arcsec for these sources. Wavelength slices in which the aperture overlapped masked mosaic edges were excluded for that source, and a star was retained only when the surviving slices still covered at least $85$ per cent of the band, so that the synthetic flux was built from a representative sampling of the bandpass. The aperture spectrum was converted from $f_\lambda$ to $f_\nu$ and folded through the appropriate ACS/WFC throughput, taken from the SVO Filter Profile Service \citep{Rodrigo2012, Rodrigo2020}, using a photon-counting band average to form the synthetic flux density. The F814W response extends beyond the red limit of the MUSE range, so its synthetic flux is built from the covered $\sim 97$ per cent of the band. After a $3\sigma$ clip, the two scales agree with a median MUSE-to-\textit{HST} ratio of $1.07$ (NMAD $0.08$, $33$ stars) in F606W and $1.05$ (NMAD $0.08$, $54$ stars) in F814W, with unity lying within the star-to-star scatter in both bands. Given the consistency with unity in both bands, we do not apply any correction for this offset to the MUSE line fluxes. A systematic reduction of the $\mathrm{Ly}\alpha$ fluxes by $5$ per cent would not significantly affect the conclusions of our analysis. The two bands sit at different central wavelengths and return consistent ratios, so we find no significant wavelength-dependent offset that would indicate an error introduced by the slice-by-slice combination and this is consistent with the mosaicking preserving the flux scale. Figure~\ref{fig:flux_cal} shows both comparisons.

\newpage
\onecolumn
\section{MUSE Spectra of Ly{\fontsize{13}{10}\selectfont$\mathbf{\alpha}$} non-detections}
\label{appendix:spectra}
\begin{figure*}
    \includegraphics[width=\textwidth]{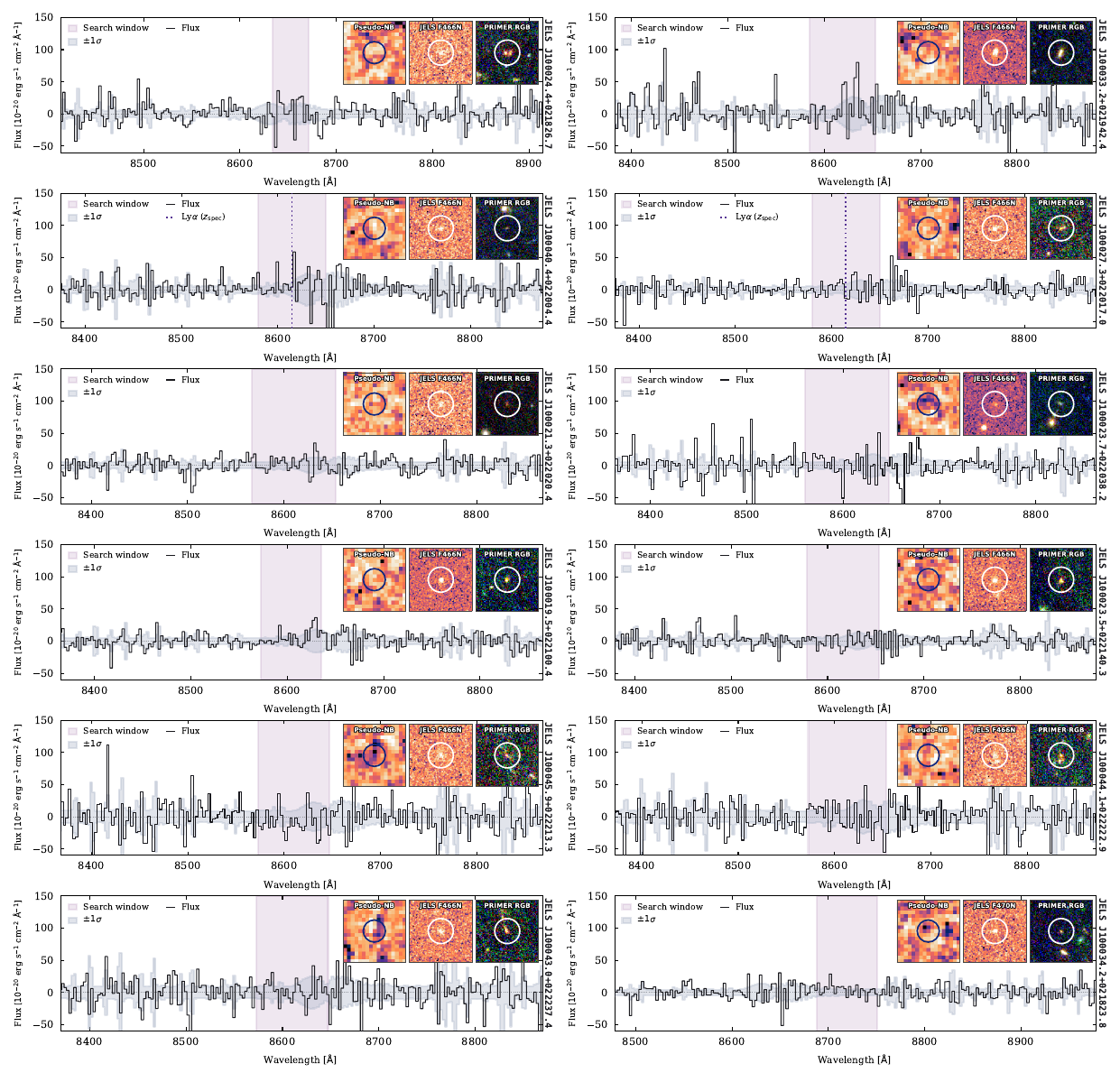}
    \caption{As Figure~\ref{fig:detect}, but for JELS H$\alpha$-selected sources with no significant Ly$\alpha$ emission detected in the MUSE spectroscopy ($\Sigma_{\mathrm{Ly}\alpha} < \Sigma_{\mathrm{Ly}\alpha}^{98}$). Upper limits on the Ly$\alpha$ flux can be inferred from the $\pm 1\sigma$ noise envelope.}
    \label{fig:nondetect}
\end{figure*}

\newpage
\onecolumn
\section{Tables of Source Properties}
\label{appendix:Tables}

\begin{table*}
\centering
\footnotesize
\setlength{\tabcolsep}{5pt}
\caption{Identification and Ly$\alpha$ properties of the H$\alpha$-selected sample at $z\simeq6.1$. Redshifts are spectroscopic where available, otherwise the MUSE Ly$\alpha$ redshift for detections and the JELS photometric redshift for non-detections. Ly$\alpha$ luminosities are given for detections and $5\sigma$ upper limits for non-detections. Physical properties for the same sources are listed in Table~\ref{tab:sample_props}. $^{\dagger}$ marks AGN candidates. \label{tab:sample_ids}}
{\renewcommand{\arraystretch}{1.5}
\begin{tabular}{lcccccc}
\hline
JELS ID & R.A. (deg) & Dec. (deg) & $z_{\rm best}$ & $f_{\rm esc}^{{\rm Ly}\alpha}$ & $L_{{\rm Ly}\alpha}$ ($10^{42}$\,erg\,s$^{-1}$) & $L_{{\rm Ly}\alpha}^{\rm lim}$ ($10^{42}$\,erg\,s$^{-1}$) \\
\hline
JELS J100038.5+021711.6$^{\dagger}$ & 150.160247 & 2.286542 & $6.0722$ & $0.002 \pm 0.001$ & $1.56 \pm 0.24$ & $\ldots$ \\
JELS J100037.2+021720.9 & 150.155101 & 2.289099 & $6.0804$ & $0.12 \pm 0.04$ & $1.60 \pm 0.46$ & $\ldots$ \\
JELS J100035.8+021806.4 & 150.149015 & 2.301786 & $6.0804$ & $0.18 \pm 0.03$ & $3.57 \pm 0.25$ & $\ldots$ \\
JELS J100024.4+021826.7 & 150.101696 & 2.307418 & $6.1186$ & $<0.30$ & $\ldots$ & $<1.28$ \\
JELS J100027.9+021920.0 & 150.116016 & 2.322179 & $6.0818$ & $0.27 \pm 0.07$ & $1.09 \pm 0.16$ & $\ldots$ \\
JELS J100033.2+021942.4 & 150.138461 & 2.328439 & $6.0928$ & $<0.15$ & $\ldots$ & $<2.13$ \\
JELS J100040.4+022004.4 & 150.168540 & 2.334553 & $6.0864$ & $<0.18$ & $\ldots$ & $<1.13$ \\
JELS J100037.6+022012.6 & 150.156806 & 2.336835 & $6.0673$ & $0.23 \pm 0.05$ & $4.81 \pm 0.19$ & $\ldots$ \\
JELS J100027.3+022017.0 & 150.113693 & 2.338062 & $6.0863$ & $<0.20$ & $\ldots$ & $<0.75$ \\
JELS J100021.3+022020.4 & 150.088831 & 2.339012 & $6.0811$ & $<0.13$ & $\ldots$ & $<0.48$ \\
JELS J100021.8+022032.8 & 150.090879 & 2.342392 & $6.0726$ & $0.83 \pm 0.26$ & $2.00 \pm 0.20$ & $\ldots$ \\
JELS J100023.7+022038.2 & 150.098843 & 2.343943 & $6.0765$ & $<0.13$ & $\ldots$ & $<0.53$ \\
JELS J100019.5+022100.4 & 150.081395 & 2.350119 & $6.0782$ & $<0.06$ & $\ldots$ & $<0.55$ \\
JELS J100018.7+022101.5$^{\dagger}$ & 150.077780 & 2.350375 & $6.0586$ & $0.04 \pm 0.01$ & $3.86 \pm 0.17$ & $\ldots$ \\
JELS J100026.3+022122.2 & 150.109598 & 2.356238 & $6.0847$ & $0.04 \pm 0.01$ & $1.01 \pm 0.24$ & $\ldots$ \\
JELS J100045.9+022213.3 & 150.191302 & 2.370349 & $6.0810$ & $<0.06$ & $\ldots$ & $<0.87$ \\
JELS J100043.0+022237.4 & 150.179232 & 2.377068 & $6.0824$ & $<0.11$ & $\ldots$ & $<1.00$ \\
JELS J100034.2+021823.8 & 150.142486 & 2.306622 & $6.1715$ & $<0.06$ & $\ldots$ & $<0.43$ \\
JELS J100039.8+021903.7 & 150.166023 & 2.317723 & $6.1581$ & $0.07 \pm 0.05$ & $1.47 \pm 0.26$ & $\ldots$ \\
JELS J100040.3+021921.5 & 150.168000 & 2.322630 & $6.1518$ & $0.38 \pm 0.09$ & $1.99 \pm 0.28$ & $\ldots$ \\
JELS J100039.7+022034.4 & 150.165262 & 2.342845 & $6.1530$ & $0.08 \pm 0.02$ & $0.90 \pm 0.14$ & $\ldots$ \\
JELS J100036.0+021823.1 & 150.150189 & 2.306428 & $6.0794$ & $0.12 \pm 0.04$ & $1.58 \pm 0.37$ & $\ldots$ \\
JELS J100023.5+022140.3 & 150.097978 & 2.360992 & $6.0854$ & $<0.03$ & $\ldots$ & $<0.59$ \\
JELS J100044.1+022222.9 & 150.183716 & 2.372950 & $6.0903$ & $<0.21$ & $\ldots$ & $<1.78$ \\
\hline
\end{tabular}
}
\end{table*}

\begin{table*}
\centering
\footnotesize
\setlength{\tabcolsep}{6pt}
\caption{Physical properties of the H$\alpha$-selected sample, corresponding to the sources in Table~\ref{tab:sample_ids} and shown in Figure~\ref{fig:esc_prop}. Percentile-based quantities are quoted with $16$th/$84$th uncertainties and the other quantities with $1\sigma$ errors. $^{\dagger}$ marks AGN candidates. \label{tab:sample_props}}
{\renewcommand{\arraystretch}{1.5}
\begin{tabular}{lcccccc}
\hline
JELS ID & $M_{\rm UV}$ & $\beta$ & EW$_{{\rm Ly}\alpha,{\rm rest}}$ (\AA) & $E(B-V)$ & $\log(M_\star/{\rm M}_\odot)$ & $\log({\rm sSFR}_{10\rm{Myr}}/{\rm yr}^{-1})$ \\
\hline
JELS J100038.5+021711.6$^{\dagger}$ & $-19.74 \pm 0.14$ & $-2.27 \pm 0.26$ & $21.2 \pm 4.4$ & $0.866^{+0.077}_{-0.085}$ & $10.79^{+0.13}_{-0.26}$ & $-8.40^{+0.22}_{-0.14}$ \\
JELS J100037.2+021720.9 & $-20.25 \pm 0.11$ & $-2.07 \pm 0.20$ & $14.2 \pm 4.4$ & $0.029^{+0.023}_{-0.012}$ & $9.22^{+0.16}_{-0.22}$ & $-8.45^{+0.43}_{-0.32}$ \\
JELS J100035.8+021806.4 & $-19.74 \pm 0.12$ & $-2.03 \pm 0.23$ & $51.1 \pm 7.2$ & $0.072^{+0.058}_{-0.033}$ & $8.42^{+0.16}_{-0.18}$ & $-7.45^{+0.23}_{-0.24}$ \\
JELS J100024.4+021826.7 & $-19.41 \pm 0.14$ & $-2.14 \pm 0.28$ & $\ldots$ & $0.043^{+0.044}_{-0.024}$ & $8.32^{+0.17}_{-0.23}$ & $-7.58^{+0.29}_{-0.27}$ \\
JELS J100027.8+021920.0 & $-18.69 \pm 0.27$ & $-2.51 \pm 0.54$ & $57.1 \pm 16.5$ & $0.019^{+0.028}_{-0.014}$ & $7.44^{+0.31}_{-0.22}$ & $-7.25^{+0.21}_{-0.37}$ \\
JELS J100033.2+021942.4 & $-20.43 \pm 0.06$ & $-1.83 \pm 0.13$ & $\ldots$ & $0.031^{+0.036}_{-0.014}$ & $9.31^{+0.14}_{-0.20}$ & $-8.44^{+0.42}_{-0.30}$ \\
JELS J100040.4+022004.4 & $-18.68 \pm 1.31$ & $-4.18 \pm 2.15$ & $\ldots$ & $0.299^{+0.256}_{-0.148}$ & $8.11^{+0.51}_{-0.39}$ & $-8.00^{+0.37}_{-0.41}$ \\
JELS J100037.6+022012.6 & $-19.25 \pm 0.16$ & $-2.20 \pm 0.31$ & $104.1 \pm 17.4$ & $0.122^{+0.073}_{-0.051}$ & $8.27^{+0.17}_{-0.26}$ & $-7.37^{+0.23}_{-0.22}$ \\
JELS J100027.3+022017.0 & $-18.81 \pm 0.22$ & $-2.93 \pm 0.43$ & $\ldots$ & $0.032^{+0.040}_{-0.022}$ & $8.00^{+0.19}_{-0.18}$ & $-7.80^{+0.24}_{-0.25}$ \\
JELS J100021.3+022020.4 & $-17.95 \pm 0.55$ & $-1.98 \pm 0.97$ & $\ldots$ & $0.080^{+0.082}_{-0.044}$ & $8.27^{+0.23}_{-0.25}$ & $-8.14^{+0.39}_{-0.42}$ \\
JELS J100021.8+022032.8 & $-19.00 \pm 0.20$ & $-3.08 \pm 0.48$ & $45.5 \pm 10.8$ & $0.013^{+0.015}_{-0.008}$ & $7.80^{+0.21}_{-0.38}$ & $-7.89^{+0.70}_{-0.49}$ \\
JELS J100023.7+022038.2 & $-18.63 \pm 0.24$ & $-1.93 \pm 0.44$ & $\ldots$ & $0.051^{+0.050}_{-0.030}$ & $8.56^{+0.19}_{-0.16}$ & $-8.28^{+0.32}_{-0.37}$ \\
JELS J100019.5+022100.4 & $-19.21 \pm 0.18$ & $-2.46 \pm 0.32$ & $\ldots$ & $0.067^{+0.057}_{-0.040}$ & $8.36^{+0.15}_{-0.17}$ & $-7.78^{+0.23}_{-0.23}$ \\
JELS J100018.7+022101.5$^{\dagger}$ & $-20.12 \pm 0.08$ & $-2.12 \pm 0.17$ & $38.3 \pm 3.6$ & $0.406^{+0.085}_{-0.083}$ & $9.60^{+0.10}_{-0.15}$ & $-7.80^{+0.22}_{-0.23}$ \\
JELS J100026.3+022122.2 & $-19.65 \pm 0.14$ & $-1.58 \pm 0.25$ & $17.4 \pm 4.8$ & $0.046^{+0.027}_{-0.013}$ & $8.75^{+0.14}_{-0.17}$ & $-7.70^{+0.26}_{-0.23}$ \\
JELS J100045.9+022213.3 & $-18.99 \pm 0.30$ & $-1.87 \pm 0.53$ & $\ldots$ & $0.098^{+0.066}_{-0.044}$ & $8.34^{+0.22}_{-0.23}$ & $-7.60^{+0.26}_{-0.33}$ \\
JELS J100043.0+022237.4 & $-19.04 \pm 0.26$ & $-3.46 \pm 0.61$ & $\ldots$ & $0.184^{+0.088}_{-0.074}$ & $8.46^{+0.17}_{-0.20}$ & $-7.73^{+0.30}_{-0.30}$ \\
JELS J100034.2+021823.8 & $-19.49 \pm 0.78$ & $-6.20 \pm 2.11$ & $\ldots$ & $0.140^{+0.141}_{-0.071}$ & $7.38^{+0.32}_{-0.24}$ & $-7.10^{+0.10}_{-0.26}$ \\
JELS J100039.8+021903.7 & $\ldots$ & $\ldots$ & $406.8 \pm 155.0$ & $0.465^{+0.261}_{-0.193}$ & $8.87^{+0.46}_{-0.38}$ & $-8.17^{+0.25}_{-0.24}$ \\
JELS J100040.3+021921.5 & $-18.40 \pm 0.35$ & $-1.61 \pm 0.62$ & $58.3 \pm 16.5$ & $0.043^{+0.049}_{-0.025}$ & $8.43^{+0.18}_{-0.24}$ & $-8.17^{+0.40}_{-0.27}$ \\
JELS J100039.7+022034.4 & $-16.66 \pm 1.88$ & $-0.06 \pm 2.88$ & $94.8 \pm 28.6$ & $0.136^{+0.085}_{-0.051}$ & $8.10^{+0.36}_{-0.31}$ & $-7.60^{+0.33}_{-0.36}$ \\
JELS J100036.0+021823.1 & $-19.42 \pm 0.32$ & $-2.52 \pm 0.89$ & $36.2 \pm 8.7$ & $0.077^{+0.071}_{-0.044}$ & $8.70^{+0.24}_{-0.21}$ & $-8.11^{+0.30}_{-0.31}$ \\
JELS J100023.5+022140.3 & $-19.60 \pm 0.49$ & $-2.67 \pm 1.58$ & $\ldots$ & $0.126^{+0.098}_{-0.062}$ & $8.68^{+0.27}_{-0.25}$ & $-7.86^{+0.28}_{-0.29}$ \\
JELS J100044.1+022222.9 & $-19.76 \pm 0.17$ & $-2.52 \pm 0.46$ & $\ldots$ & $0.035^{+0.043}_{-0.022}$ & $8.85^{+0.17}_{-0.19}$ & $-8.30^{+0.34}_{-0.36}$ \\
\hline
\end{tabular}
}
\end{table*}


\bsp	
\label{lastpage}
\end{document}